\documentclass[twocolumn,appendixfloats]{aastex62}

\usepackage{hyperref}
\usepackage{amsmath,amstext}
\usepackage[T1]{fontenc}
\usepackage{apjfonts} 
\usepackage[figure,figure*]{hypcap}
\usepackage{gensymb}
\usepackage{xcolor}
\usepackage{soul}
\usepackage[normalem]{ulem}

\newcommand{\gmos}{\emph{Gemini}/GMOS}
\newcommand{\imacs}{\emph{Magellan}/IMACS}

\newcommand{\tdg}{\tablenotemark{$\dagger$}}
\newcommand{\tddg}{\tablenotemark{$\ddagger$}}

\def \wf {0.9} 

\shorttitle{ACCESS: Ground-based Optical Transmission Spectroscopy of the Hot Jupiter WASP-4b}
\shortauthors{Bixel et al.}

\begin{document}
\title{ACCESS: Ground-based Optical Transmission Spectroscopy of the Hot Jupiter WASP-4b}

\author{Alex Bixel}
\affiliation{Department of Astronomy/Steward Observatory, The University of Arizona, 933 N. Cherry Avenue, Tucson, AZ 85721, USA}
\affiliation{Earths in Other Solar Systems Team, NASA Nexus for Exoplanet System Science}
\email{abixel@email.arizona.edu}

\author{Benjamin V. Rackham}
\affiliation{Department of Astronomy/Steward Observatory, The University of Arizona, 933 N. Cherry Avenue, Tucson, AZ 85721, USA}
\affiliation{Earths in Other Solar Systems Team, NASA Nexus for Exoplanet System Science}

\author{D\'aniel Apai}
\affiliation{Department of Astronomy/Steward Observatory, The University of Arizona, 933 N. Cherry Avenue, Tucson, AZ 85721, USA}
\affiliation{Earths in Other Solar Systems Team, NASA Nexus for Exoplanet System Science}
\affiliation{Lunar and Planetary Laboratory, The University of Arizona, 1640 E. University Blvd, AZ 85721, USA}

\author{N\'estor Espinoza}
\affiliation{Max-Planck-Institut f\"ur Astronomie, K\"onigstuhl 17, 69117, Heidelberg, Germany}
\affiliation{Instituto de Astrof\'isica, Facultad de F\'isica, Pontificia Universidad Cat\'olica de Chile, Av. Vicu\~na Mackenna 4860, 782-0436 Macul, Santiago, Chile}
\affiliation{Millennium Institute of Astrophysics, Av. Vicu\~na Mackenna 4860, 782-0436 Macul, Santiago, Chile}

\author{Mercedes L\'opez-Morales}
\affiliation{Harvard-Smithsonian Center for Astrophysics, 60 Garden Street, Cambridge, MA 01238, USA}
\affiliation{Earths in Other Solar Systems Team, NASA Nexus for Exoplanet System Science}

\author{David Osip}
\affiliation{Las Campanas Observatory, Carnegie Institution of Washington, Colina el Pino, Casilla 601, La Serena, Chile}

\author{Andr\'es Jord\'an}
\affiliation{Instituto de Astrof\'isica, Facultad de F\'isica, Pontificia Universidad Cat\'olica de Chile, Av. Vicu\~na Mackenna 4860, 782-0436 Macul, Santiago, Chile}
\affiliation{Millennium Institute of Astrophysics, Av. Vicu\~na Mackenna 4860, 782-0436 Macul, Santiago, Chile}
\affiliation{Max-Planck-Institut f\"ur Astronomie, K\"onigstuhl 17, 69117, Heidelberg, Germany}
\affiliation{Earths in Other Solar Systems Team, NASA Nexus for Exoplanet System Science}

\author{Chima McGruder}
\affiliation{Harvard-Smithsonian Center for Astrophysics, 60 Garden Street, Cambridge, MA 01238, USA}

\author{Ian Weaver}
\affiliation{Harvard-Smithsonian Center for Astrophysics, 60 Garden Street, Cambridge, MA 01238, USA}

\begin{abstract}

We present an optical transmission spectrum of the atmosphere of WASP-4b obtained through observations of four transits with \imacs. Using a Bayesian approach to atmospheric retrieval, we find no evidence for scattering or absorption features in our transit spectrum. Our models include a component to model the transit light source effect (spectral contamination from unocculted spots on the stellar photosphere), which we show can have a marked impact on the observed transmission spectrum for reasonable spot covering fractions ($<5\%$); this is the first such analysis for WASP-4b. We are also able to fit for the size and temperature contrast of spots observed during the second and third transits, finding evidence for both small, cool and large, warm spot-like features on the photosphere. Finally, we compare our results to those published by \citet{Huitson17} using \gmos\ and \citet{May18} using IMACS, and find that our data are in agreement.
\end{abstract} 


\section{Introduction} \label{sec:introduction}
\object{WASP-4b} is a $1.4\,R_J$ hot Jupiter orbiting a G7V star with a period of 1.34 d, equilibrium temperature of $\sim 1700$ K, and transit depth $(R_p/R_s)^2 = 2.4\%$ \citep{Wilson08}. It has been observed in transit over three dozen times, offering strong constraints on its orbit \citep{Hoyer13}, the spot activity and relative rotation of its host star \citep{Sanchis-Ojeda11}, and also placing upper limits on its transit timing variation amplitude \citep{Nikolov12}.

\cite{Beerer11} used \emph{Spitzer} to observe the planet's secondary eclipse and place constraints on the temperature profile of its atmosphere, and they conclude that the evidence is consistent with either a weak temperature inversion or none at all. However, even stronger detections of thermal inversions using \emph{Spitzer} have later been called into question \citep[e.g., HD 209458b, ][]{Diamond-Lowe14}. The evidence for an inversion therefore remains marginal, leaving us with little insight about the composition of the planet's upper atmosphere.

Given the tight constraints on its transit depth and orbital properties, WASP-4b is a natural target for spectroscopic transit observations. Transit spectroscopy can be used to constrain the composition and structure of the planet's upper atmosphere and test for the presence of clouds, scattering hazes, and atomic or molecular absorbers \citep[e.g.,][]{Seager00a,Brown01,Hubbard01}. These observations also offer insight into the stellar photosphere, i.e., through the measurement of star spot temperatures within the transit chord \citep[e.g.,][]{Pont08, Sing11, Beky14}. However, signals from the photosphere can be degenerate with those from the planet's atmosphere, leading to contrasting interpretations of planetary and stellar origins for features in optical transmission spectra \citep[e.g.,][]{Pont13, McCullough14}. It is therefore critical to demonstrate methods for accounting for this degeneracy as the field moves toward smaller targets and increasingly precise observations \citep[e.g.,][]{Rackham18}.

Transit spectroscopy of WASP-4b has been attempted with \emph{HST}/WFC3 \citep{Ranjan14}, but was unsuccessful due to detector saturation. A \gmos\ optical transmission spectrum of WASP-4b has been published by \citet{Huitson17}, who measure a nearly uniform opacity from 440-940 nm, suggesting the presence of high-altitude clouds and a possible sodium absorption feature. More recently, \cite{May18} have published a transmission spectrum using \imacs, and have similarly found no evidence for spectral features.

As part of the Arizona-CfA-Cat\'olica-Carnegie Exoplanet Spectroscopy Survey \citep[ACCESS,][]{Rackham17}, we have observed four transits of WASP-4b with \imacs. We have previously demonstrated the use of this instrument for transit spectroscopy using our custom data reduction pipeline \citep{Jordan13,Rackham17,Espinoza19}. In this paper, we present an optical transmission spectrum from 450-900 nm. We interpret our results using a Bayesian retrieval code introduced in \cite{Espinoza19} and find no evidence for scattering or absorption features. We also fit for the size and temperature of photosphere features occulted during the second and third transits, and derive corrections for their effects on the transit spectrum. Finally, we compare our findings with those of \cite{Huitson17} and \cite{May18}.

\section{Observations} \label{sec:observations}
We conducted spectroscopic observations of WASP-4 on the nights of 24 September 2013, 17 October 2013, 14 August 2015, and 26 September 2015 (hereafter Transits 1--4) using the Inamori-Magellan Areal Camera \& Spectrograph \citep[IMACS,][]{Dressler11} on the 6.5m Magellan-Baade telescope at Las Campanas Observatory in Chile. We observed using multi-slit masks in the f/2 mode with 2x2 binning (0.4"/px). Observations of HeNeAr and quartz calibration lamps before and after the observations allowed for wavelength calibration and flat-field correction. The key parameters of our observations are listed in Table \ref{tab:transit_data}.

\subsection{24 September 2013, 17 October 2013, and 14 August 2015}
On the first three nights we used a setup consisting of a 400--1000~nm spectroscopic filter, a 300 lines/mm grism with a blaze angle of 17.5 degrees, and 10" wide by 20" long spectral slits for the target and reference stars. Twelve reference stars were observed, although only one was used in the final data reduction for reasons discussed in Section \ref{sec:reference_stars}. Most of the stellar spectra were dispersed across two chips. While the lunar sky background was minimal on the first and third nights, it was substantial during the second, and we take this into account in our data reduction pipeline. Finally, the observations on 24 September 2013 (Transit 1) did not commence until shortly after ingress, so we were only able to observe a partial transit.

\subsection{26 September 2015}
Our instrument setup for the fourth observation was modified from the previous three. We used a 570--980~nm order-blocking filter for the purpose of eliminating higher-order interference towards red wavelengths, 10" wide by 10" long slits, and a 150 lines/mm grism with a blaze angle of 18.8 degrees. This setup allowed for more tightly dispersed spectra which fall on a single chip, thereby reducing detector-to-detector variations in the spectra and avoiding chip gaps. The moon was in full phase and separated from the target by 40 degrees, contributing significantly to the ambient sky background, which we again account for in our data reduction pipeline.

\begin{deluxetable*}{ccccccc}
\tablecaption{Instrument setup and transit model characteristics for each of our four observations. All observations made use of the IMACS f/2 camera with multi-slit masks. The ratio in column 7 factors in the brightness of the target and single reference star, but not the sky background. \label{tab:transit_data}}
\tablehead{Transit & Date (start of night) & Filter & Grism (l/mm) & Airmass & $R_p/R_s$ (white light) & Scatter / Photon noise}
\startdata
1 & 24 September 2013 	& Spectroscopic f/2	& 300		& < 1.3	& $0.1528^{+0.0012}_{-0.0011}$	& 4.9 \\
2 & 17 October 2013   	& \ldots 			& \ldots 	& < 1.1	& $0.1537^{+0.0008}_{-0.0007}$	& 7.8 \\
3 & 14 August 2015    	& \ldots 			& \ldots 	& < 1.7	& $0.1544^{+0.0002}_{-0.0002}$	& 2.9 \\
4 & 26 September 2015	& WBP 5694-9819		& 150		& < 1.2	& $0.1565^{+0.0004}_{-0.0005}$	& 6.2 \\
\enddata
\end{deluxetable*}

\section{Data reduction} \label{sec:data_reduction}

We reduce the raw data using our custom Python-based pipeline which has been used for similar observations of \object{WASP-6b} \citep{Jordan13}, \object{GJ~1214b} \citep{Rackham17}, and \object{WASP-19b} \citep{Espinoza19}. In the following paragraphs, we give a brief overview of the pipeline functions. A more detailed review can be found in \citet{Jordan13} and \citet{Rackham17}; the only more recent addition to the pipeline is the correction for the non-uniform sky background described in Section \ref{sec:lbr}.

We use quartz lamp images taken with the same configuration as the science images to apply a flat-field correction, and we calculate the bias offset from the overscan region of each chip. We use full-frame flats taken without a mask or grism to identify bad pixels, and we identify cosmic rays on the stellar spectra using a $3\sigma$ threshold at each row along the direction of spectral dispersion. To trace the stellar spectra, we calculate the centroid in each row along the dispersion direction and fit a second-order polynomial to the centroid values. For Transits 1 and 2, we assume the background is uniform in the spatial direction, measure it using the outermost 14 px in each row along the dispersion direction, and subtract. For Transits 3 and 4, the background is not spatially uniform, so we subtract it using the method outlined in Section \ref{sec:lbr}. Finally, to extract the spectra we use the optimal extraction algorithm outlined by \cite{Marsh89}, which involves fitting a third-order polynomial to the spectral profile at each row along the dispersion direction, then using that profile to weight each pixel when summing the flux.

We use HeNeAr arc lamp exposures taken before and after the transit observations to calculate the wavelength solutions for each star, which convert the pixel coordinates along the dispersion direction to wavelength values. We manually identify prominent emission lines and use a sixth-order polynomial to fit the wavelength solution; given the large number of lines used, the danger of over-fitting is minimal. The marked lines are iteratively rejected based on their residuals to the fit until the residuals are below $\sim 0.05$ \AA.

Our spectra drift in the dispersion direction over the course of the night, resulting in frame-to-frame offsets  in the wavelength solution. To solve this, we cross-correlate the first spectrum of a star with each of the subsequent spectra to determine the shift in wavelength, then fit a third-order polynomial to this shift as a function of time and use it to correct the wavelength solution in each frame.

The output of the pipeline is a set of reduced wavelength-binned and integrated (``white'') light curves for the target and reference stars.

\subsection{Non-uniform sky background} \label{sec:lbr}
In some of our IMACS data sets for this and other ACCESS targets, we have noticed that spatially uniform sources of light inherit a non-uniform profile in the spatial direction once the mask and dispersive element are in place. This effect also widens the spectra of point sources. We have attributed this to internal scattering within the instrument, and have found that it is more common in newer images: of the four data sets for WASP-4, only Transits 3 and 4 are affected. We measure the extent of the effect using our flat field images, finding that the profile of spatially uniform light peaks at the slit center and appears $\sim 10\%$ fainter at the slit edges. Unless we account for this, we will underestimate the sky background when we extract the stellar spectra.

For the two affected data sets, we use quartz lamp exposures taken at the beginning of the night to model the scattered light profile. The peaked profiles are not consistent with common symmetrical functions (e.g., Gaussian), nor are they well-described by a classical high-order polynomial, because high-order polynomials commonly fail at the edges of their fitted intervals (an effect known as Runge's phenomenon). Instead, we use Chebyshev polynomials which are not as susceptible to this effect, and find that sixth-order polynomials are sufficient to match the data.

The polynomials are fitted independently for every row along the dispersion direction of every slit in the flat field images. Then, in each science image we fit the amplitudes of the polynomials using the background level in the outermost 14 px in each row along the dispersion direction (8 px for Transit 4 due to shorter slits). Finally, we subtract the fitted polynomial from each row.

\subsection{Flat field correction}
Applying the flat-field correction for Transits 1, 3, or 4 does not significantly change our binned light curves, or the transit spectra which we derive in Section \ref{sec:transit_fit}. However, applying the correction for Transit 2 introduces $\sim 2\%$-level, non-linear time-dependent trends into out-of-transit baseline flux of the binned light curves blueward of 530 nm. These trends remain even when normalizing by the reference star as discussed in the following. Therefore, for consistency, we choose not to apply the flat-field correction to any of our data sets. 

\subsection{Reference star selection} \label{sec:reference_stars}
The shape of the target's light curve is complicated by instrumental and atmospheric effects, such as changes in airmass or transparency. To calibrate out these effects, we simultaneously observed 12 reference stars of comparable optical apparent magnitudes and color ratios using multi-slit masks. Of these, two spectra reached the saturation limit of the detector and were not usable.

We use our highest quality data set (Transit 3) to determine which of the ten remaining reference stars to use in our light curve analysis. Our primary consideration is the shape of the out-of-transit baseline for the target light curve when normalized by each star. Eight of the stars leave residual long-term trends at the $\sim 1\%$ level. Two stars leave trends at the $\sim 0.1\%$ level, and they happen to be the only two stars occupying the same pair of detector chips as the target. The point-to-point scatters of the baseline points for the two resulting light curves are 0.4 and 0.6 mmag, which are both smaller than the scatters for the eight light curves with larger long-term trends; we therefore further restrict our set of potential reference stars to these two. We perform a similar analysis of the two remaining stars using our three other data sets, and discover that one of the two stars reliably produces a flat baseline, while the other star introduces trends at the $\sim 0.5-1\%$ level. Therefore, we discard this star as well.

We are left with a single reference star, \object{2MASS J23341836-4204509}, which reliably produces a flat out-of-transit baseline with a low point-to-point scatter in all of our data sets. We note that this reference star was also used in a similar analysis by \cite{Huitson17}. The spectral type of the reference star has not been previously published, but using parallax measurements from Gaia DR2 \citep{Gaia2018} we calculate that it is intrinsically brighter than the target by $\sim 80\%$ from 400 - 1000 nm. While this wavelength range does not capture all of the emitted light, it does cover the emission peak for late F and all G stars. Using a simple $L_* \propto M_*^{3.5}$ scaling relation, and assuming that our wavelength range covers most of the emitted light, this suggests that the reference star is $\sim 20\%$ more massive than the G7V, 0.85 $M_\odot$ target \citep{Gillon09}, consistent with an early G or late F dwarf.

\subsection{Effects of atmospheric dispersion}\label{sec:stability}
Magellan-Baade is equipped with an atmospheric dispersion compensator (ADC), which corrects for the effects of differential atmospheric refraction as the airmass of our target changes over the course of the night. If left uncorrected, differential refraction would affect our stellar wavelength solutions in a non-linear manner, ``stretching'' or ``shrinking'' the spectra over the course of the night. We test the accuracy of the ADC correction by measuring the distance in calibrated wavelength space between Na and H$\alpha$ in the target and reference spectra as a function of time, which we calculate to vary by no more than 1 \AA\ over the course of the night. Furthermore, the difference between the measured distance for the target and reference stellar spectra changes by no more than 0.5 \AA, and less than 0.1 \AA\ on nights with higher quality data. Since the long-term change in this measurement is much lower than the frame-to-frame scatter (up to 5 \AA), we reason that any effect on the measured transit depth will be negligible.

\subsection{Observing efficiency}
During Transits 1 and 2, we opted for longer, low-noise readout modes and short exposure times to avoid detector non-linearity. As a result, we spent only $\sim 20\%$ of our time gathering light. By comparison, we achieved an efficiency of $\sim 70\%$ and $\sim 50\%$ during Transits 3 and 4, respectively, as we chose the fastest readout mode and took longer exposures. The difference in read noise between the fastest (5.6 e-) and slowest (2.8 e-) modes is negligible compared to the Poisson noise ($>$100 e-), and as we show in Section \ref{sec:linearity}, the detector is sufficiently linear for our purposes provided the pixel values remain near or below half-well.

Based on this experience, we advise that observers prioritize longer exposure times with fast readouts for transit spectroscopy with IMACS.

\subsection{Sky background}

Transits 2 and 4 suffer from high sky background values due to the target's proximity to the full-phase moon. The effects can be measured by comparing the scatter of the residuals in Figure \ref{fig:lightcurves}. Scaled for exposure time, the magnitudes of the scatter for Transits 1 and 3 agree to within $15\%$, while the scatter of Transit 2 is $\sim 50\%$ larger. The scatter of Transit 4 is similarly $\sim 50\%$ larger, although Transit 4 was observed in a redder filter with less sky contamination.

\subsection{Slit losses} \label{sec:slit_losses}

We consider whether slit losses may have been significant during our observations. For Transits 1-3, the slits were 10" wide and 20" long, while for Transit 4 the slits were 10" wide and 10" long. We perform a least-squares fit of a Moffat profile to the point spread function of a bright but unsaturated star in each of the field acquisition images, then integrate to determine how much light would lie outside of the slit if placed over the star. For every night, the fraction of light lost is more than an order of magnitude lower than the Poisson noise.

\subsection{Detector linearity}\label{sec:linearity}
We observed Transits 3 and 4 with longer integration times to reduce the relative readout overhead and improve our signal-to-noise ratios. However, this increased the maximum pixel values of the target spectrum from $\sim 15,000$ to $\sim 30,000$ ADU, nearly half of the full well value ($\sim 65,000$ ADU). This raises the question of whether the detector's response could have become non-linear (i.e. the gain was not uniform over the range of pixel values), which could bias the mean-subtracted transit spectra which we extract later in Section \ref{sec:transit_fit}.

In principle, non-linearity should not be an issue for measuring the relatively small changes in transit depth from bin to bin, as the gain should not be expected to change over the small range of changing values. However, the pixel values at the peak of the target spectrum varied by up $15\%$ from frame to frame and $40\%$ over the course of each night, due mostly to seeing variations which changed the point spread function. Therefore it is worthwhile to determine the threshold beyond which non-linearity could significantly affect our results.

\begin{figure}
    \centering
    \includegraphics[width=0.5\textwidth]{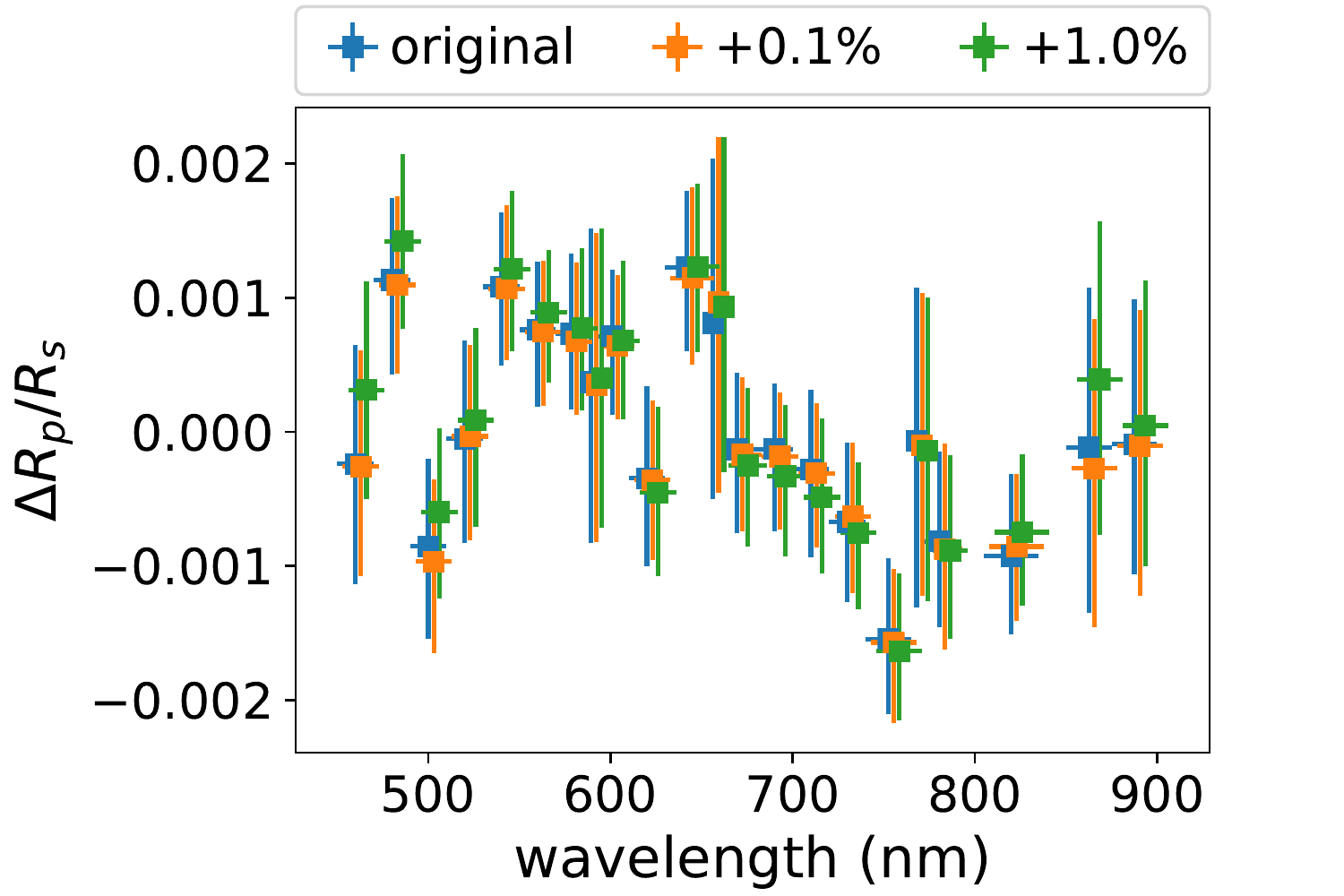}
    \caption{The mean-subtracted transit spectrum of Transit 3 with different gain modifications. We calculate and apply a gain enhancement to each pixel in the raw data, assuming the gain increases linearly by 0\%, 0.1\%, or 1\% from 0-35,000 ADU. Provided the detector is linear to $<1\%$ over this range, the mean-subtracted transit spectrum is largely unaffected.\label{fig:linearity}}
\end{figure}

To do so, we modify the raw data for Transit 3 to reflect a 0.1\% or 1\% linear increase in the gain over the range of 0-35,000 ADU. We then reduce the modified data and extract the mean-subtracted transit spectra for both gain prescriptions, which we plot alongside the original spectrum in Figure \ref{fig:linearity}. We find an average effect of $0.07 \sigma$ per bin for a 0.1\% increase in gain, and $0.23 \sigma$ per bin for a 1\% increase.

So long as the detector is linear to 0.1\% from zero to half-well, non-linear effects should have a negligible impact on our results. Even in the case of 1\% non-linearity, the effect is $\ll 1 \sigma$ except in a few bins. We are therefore confident that our results are robust to realistic levels of non-linearity, and we advise that future observers target a similar range of peak pixel values ($\sim 35,000$ ADU) so as to optimally balance observational efficiency and detector linearity.

\section{Light curve modeling} \label{sec:transit_fit} 
We model our integrated and spectroscopic light curves using the analytic models introduced by \cite{Mandel02}, and marginalize over the parameter space with a Markov-Chain Monte Carlo (MCMC) algorithm previously detailed in \citet{Rackham17}. As in previous works, we fit for the limb-darkening coefficients in order to account for any biases that might arise due to our imperfect knowledge of the intensity profile of the photosphere \citep{Espinoza15}. As WASP-4b has been studied extensively through photometric observations, we hold the orbital parameters fixed to the mean values in Table \ref{tab:properties}.

The fitted parameters for the transit model include the planet-to-star radius ratio ($R_p/R_s$), two parameters for a quadratic limb-darkening law which are sampled according to the method detailed in \cite{Kipping13}, and the mid-transit time. The baseline flux is modeled as a second-order polynomial; this decision is discussed in more detail in Section \ref{sec:systematics}.

We use the likelihood function of \cite{Carter09}, Equation 41, in which the noise in the light curve is parameterized by two free noise parameters, $\sigma_w$ and $\sigma_r$. $\sigma_w$ describes the amplitude of uncorrelated (``white'') sources of noise (e.g. photon noise), while $\sigma_r$ describes the amplitude of the correlated (``red'') noise, which is modeled as a superposition of time-localized oscillating signals known as wavelets. The power spectral density of the red noise is modeled as $S(f) \propto 1/f^\gamma$; following the authors' example, we set $\gamma = 1$. We calculate the wavelet functions and likelihoods using our own Python module\footnote{\url{https://github.com/nespinoza/flicker-noise}}.

We first fit a model to the white light curve of each transit to determine the mid-transit time, then fit each binned light curve independently with the mid-transit time fixed. The fitted white light curves for each night are shown in Figure \ref{fig:lightcurves}, and the fitted binned light curves are presented in Appendix Figures~\ref{fig:130925} through \ref{fig:150927}.

Note that two of the transits exhibit spot-crossing features which have been excluded from the fit. To determine which data points to exclude, we first conservatively remove data near the spot and fit a model to the white light curve, then fit a one-dimensional Gaussian model to the shape of the spot in the residuals, and finally exclude only those data points which lie within two standard deviations of the fitted mean. For a more detailed analysis of these spot features, see Section \ref{sec:starspots}.

The primary output of our fitting routine is a transmission spectrum for 19 independently fitted continuum bins, each approximately 20 nm in width, and three narrower bins centered on possible absorption features, including Na D, H$\alpha$, and the K I doublet (767/770 nm). We select the bin width to be as small as possible while maintaining signal-to-noise ratios of at least a few hundred in each bin and data set. We attempt to keep the bin sizes consistent across the spectrum, but some bins have been adjusted to accommodate the three narrow bins and two chip gaps. The spectra for each transit are presented in Figure~\ref{fig:specs_all}, covering wavelengths from 450--900~nm in the first three nights, and 570--900~nm for the fourth night. We have no data from 790-805 nm and 835-850 nm as these correspond to the location of the detector chip gaps in the target and reference star spectra.

\subsection{Combining data from separate nights}
The average values of $R_p/R_s$ in the spectrum of Transits 1-3 agree to within $1.5\sigma$, but Transit 4 disagrees at the $\sim 5 \sigma$ level. This mirrors the deviation in the fitted orbital parameters for Transit 4 discussed in Section \ref{sec:orbit}, and the causes are likely the same. 

Regardless of the reason for this disagreement, we are primarily interested in the \emph{change} in $R_p/R_s$ with wavelength. We subtract the weighted mean $\overline{R_p/R_s}$ from the individual spectra for Transits 2-4, then calculate the average values of $\Delta R_p/R_s$ in each bin, weighted by the uncertainties for each night. The combined transmission spectrum is presented in Figure \ref{fig:spec_combined}.

Due to the large uncertainties and incomplete baseline coverage for Transit 1, we do not include its spectrum in the combined result; see Section \ref{sec:phase_coverage}. We also correct the spectrum of Transit 3 for known stellar contamination before incorporating it into the combined spectrum, as detailed in Section \ref{sec:spot_correction}.

\subsection{Excluded bins}\label{sec:excluded_bins}

Two of the fitted bins have been excluded from our primary figures and results: the 450-470 nm bin of Transit 2, and the 875-900 nm bin of Transit 4. The values for both bins deviate by $> 3\sigma$ from the average value of their respective spectra, and both bins are at the low S/N ends of the stellar spectrum. For reference, these bins are included in Appendix Figures \ref{fig:131018} and \ref{fig:150927} and Table \ref{tab:data}.

The 450-470 nm bin of Transit 2 lies $\sim 3\sigma$ \emph{below} the average value of the spectrum; we can think of no astrophysical explanation for this effect. Furthermore, the likelihood of a statistical $3\sigma$ outlier across all of our 82 bins is $\sim 0.3\%$. Most likely, our polynomial model is inadequate to describe the low S/N systematic trends in this bin.

The 875-900 nm bin of Transit 4 lies $\sim 4\sigma$ above the weighted mean. While this wavelength range does correspond with water absorption, such an offset is not observed in the other spectra. Furthermore, the value of this bin depends strongly on whether a linear or quadratic polynomial is used to model the statistics, as discussed in Section \ref{sec:systematics}. Again, it seems most likely that our systematics model is inadequate for this low S/N bin.

\begin{deluxetable}{llc}
\tablecaption{Relevant previously measured properties of WASP-4 and its companion, with $1\sigma$ uncertainties.\label{tab:properties}}
\tablehead{\colhead{Parameter}&\colhead{Value}&\colhead{Reference}}
\startdata
WASP-4 				& 									& \\ \hline
$R_s$ $(R_\odot)$ 	& $0.873^{+0.036}_{-0.027}$			& \citet{Gillon09} \\
$M_s$ $(M_\odot)$   & $0.85^{+0.11}_{-0.07}$            & \ldots \\
{[Fe/H]} 			& $-0.03^{+0.09}_{-0.09}$			& \ldots \\
log(g) (cgs)		& $4.487^{+0.019}_{-0.015}$ 		& \ldots \\ 
$T_s$ (K)			& $5540^{+55}_{-55}$				& \citet{Maxted11} \\ \hline
WASP-4b 			& 									& \\ \hline
P (d)				& $1.33823204$	 					& \citet{Hoyer13} \\
$R_p/R_s$ 			& $0.15445^{+0.00025}_{-0.00025}$ 	& \ldots \\
$i$ (deg)			& $88.52^{+0.39}_{-0.26}$			& \ldots \\
$a/R_s$				& $5.463^{+0.025}_{-0.020}$			& \ldots \\
$e$					& $\approx 0$						& \citet{Beerer11} \\
$R_p$ $(R_J)$ 		& $1.395^{+0.022}_{-0.022}$  		& \citet{Hoyer13} \\
$M_p$ $(M_J)$ 		& $1.237^{+0.021}_{-0.021}$  		& \citet{Winn09} \\
\enddata
\end{deluxetable}

\subsection{Fitting orbital parameters} \label{sec:orbit}
We also use our MCMC code to simultaneously fit for the transit depth, the inclination ($i$) and the semi-major axis ($a/R_s$), while keeping the eccentricity and period fixed. The results of these fits are listed in Table \ref{tab:orbit}. The parameters from Transits 1-3 are generally consistent with those measured by \cite{Hoyer13}, although the values from the partial Transit 1 are poorly constrained.

The parameters from Transit 4 deviate $2-3\sigma$ from the previously measured values. It is not clear why this is the case. Stellar variability may be a partial contributor; as discussed in Section \ref{sec:heterogeneity}, WASP-4 is variable at the 6 mmag level, and this can lead to changes in the apparent transit depth. The deviation may also be related to losses due to the shorter slits employed during Transit 4, as discussed in Section \ref{sec:slit_losses}.

Due to this this large deviation, we also try fitting the spectrum of Transit 4 by fixing the inclination and semi-major axis to the fitted values, rather than those in the literature. The result is only a $\ll 1 \sigma$ change in the binned values for $\Delta R_p/R_s$, so we believe the spectrum for this night is robust despite the disagreement in the orbital parameters. For consistency with the other data sets, we keep the parameters fixed to the literature values in the remaining sections.

\begin{deluxetable}{lccccc}
\tablecaption{Fitted orbital parameters when modelling the planet radius, inclination, and semi-major axis jointly. For comparison we include the values derived by \cite{Hoyer13}. \label{tab:orbit}}
\tablehead{\colhead{}&\colhead{$R_p/R_s$}&\colhead{$a/R_s$}&\colhead{$i$ (deg)}}
\startdata
Hoyer+13       & $0.15445_{-0.00025}^{+0.00025}$ & $5.463_{-0.020}^{+0.025}$ & $88.52_{-0.26}^{+0.39}$\\\hline
Transit 1 & $0.15386_{-0.00174}^{+0.00135}$ & $5.386_{-0.157}^{+0.088}$ & $89.33_{-0.73}^{+0.47}$\\
Transit 2 & $0.15617_{-0.00167}^{+0.00164}$ & $5.439_{-0.092}^{+0.070}$ & $87.90_{-0.85}^{+1.24}$\\
Transit 3 & $0.15388_{-0.00047}^{+0.00045}$ & $5.473_{-0.038}^{+0.021}$ & $88.94_{-0.52}^{+0.53}$\\
Transit 4 & $0.15654_{-0.00065}^{+0.00067}$ & $5.520_{-0.030}^{+0.026}$ & $89.46_{-0.60}^{+0.37}$\\
\enddata
\end{deluxetable}

\begin{figure*}[p]
\centering
\includegraphics[width=\wf\textwidth]{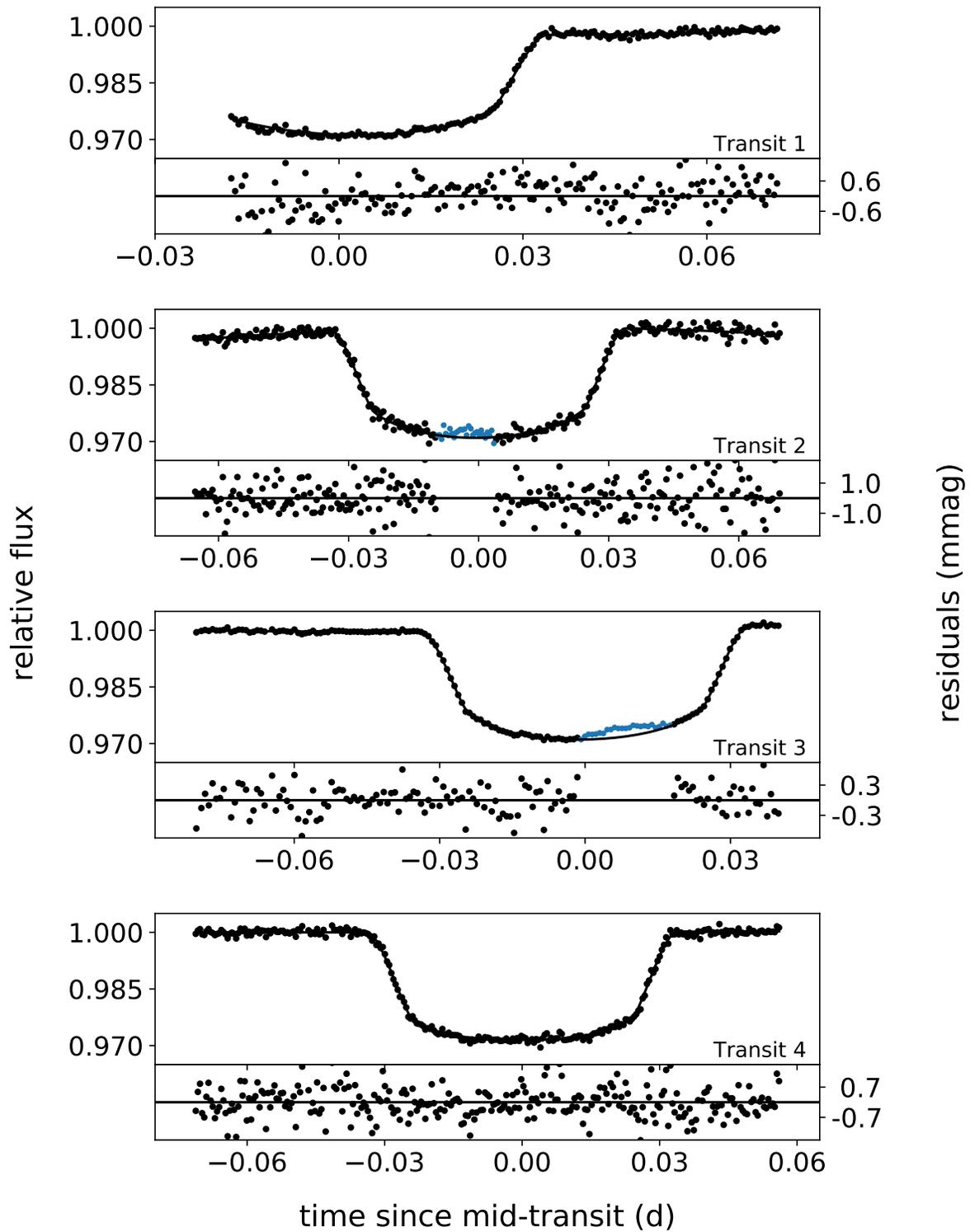}
\caption{White light curves with transit models and residuals ($1\sigma$ values marked). Two of the light curves featured spot-crossing events (blue) which are not included in the fit. The long-term trend is modeled by a second-order polynomial. \label{fig:lightcurves}}
\end{figure*}

\begin{figure*}[p]
\centering
\includegraphics[width=\wf\textwidth]{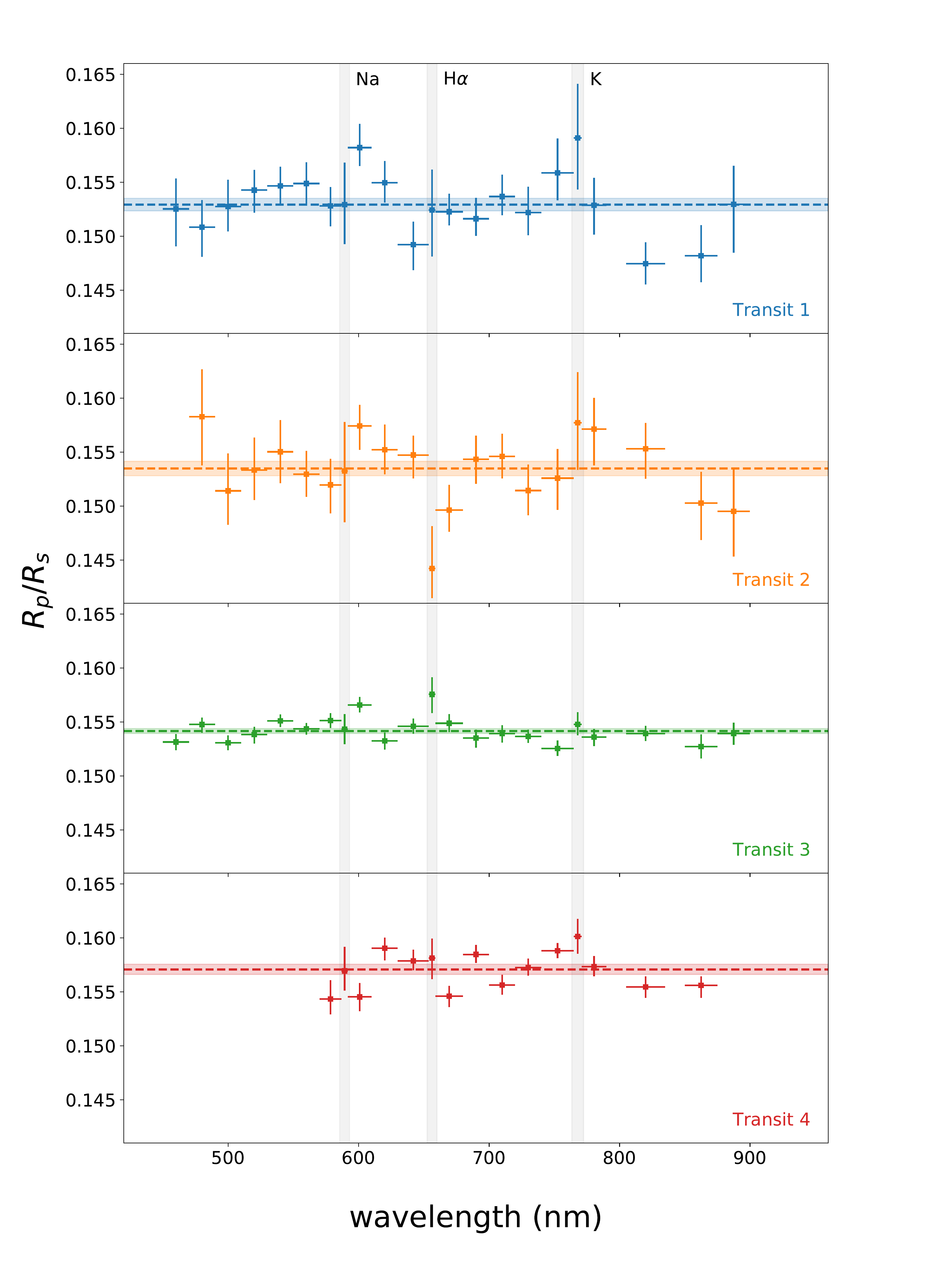}
\caption{Transmission spectra from each night, with no corrections or offsets applied. The dashed lines and shaded regions represent weighted average values with $\pm 1 \sigma$ uncertainties. The wavelengths of potential atomic features are highlighted. \label{fig:specs_all}}
\end{figure*}

\begin{figure*}[ht]
\centering
\includegraphics[width=\wf\textwidth]{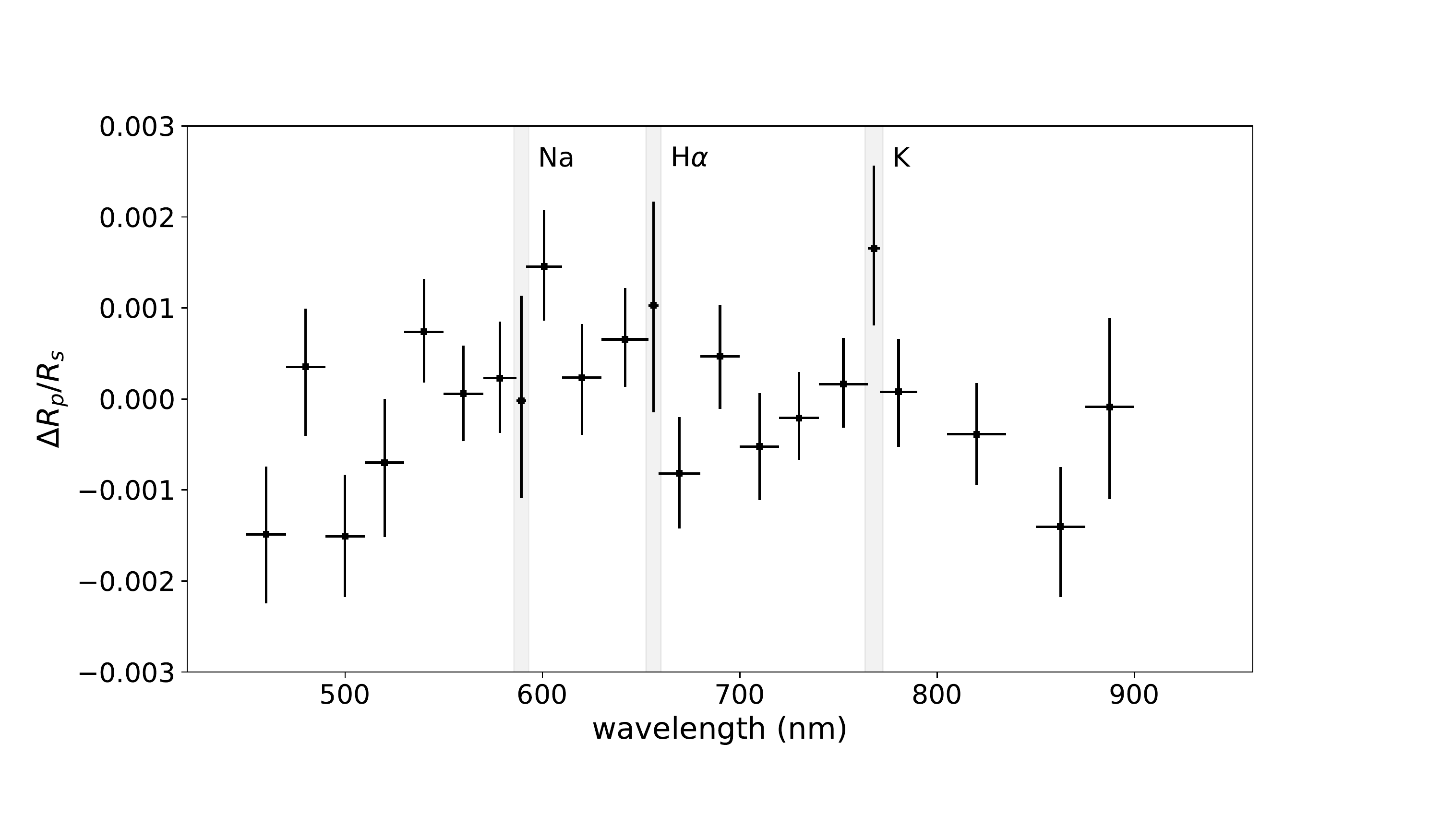}
\caption{Combined transmission spectrum from Transits 2-4, with the weighted average value subtracted from each night. Narrow bins centered on potential atomic features are highlighted, and the remaining (continuum) bins are black. The spectrum of Transit 3 is corrected for the effect of an occulted spot as described in Section \ref{sec:spot_correction} before it is incorporated into the combined result. Transit 1 is excluded due to poor phase coverage. \label{fig:spec_combined}}
\end{figure*}

\subsection{Systematics model} \label{sec:systematics}
We find that the trends which remain in the target light curves after dividing by the reference star light curves correlate with airmass, which tends to reach its minimum near the center of the transit. Nevertheless, no physically motivated (e.g., exponential) function of airmass is able to describe this trend better than a simple second-order polynomial in time, so we opt to use time-dependent polynomials to characterize our long-term systematics.

The long-term trends in many of our binned light curves cannot be adequately described by a linear polynomial in time, so we use a quadratic polynomial instead. In principle, any degeneracy between the transit depth and the polynomial value should be incorporated into our MCMC uncertainties. To measure the impact of the model choice on our results, we apply both linear and quadratic systematics models and compare the values of $\Delta R_p/R_s$ for each.

For Transits 1-3, the average absolute change in $\Delta R_p/R_s$ between the linear and quadratic models is approximately $0.5\sigma$ per bin, and no larger than $1.6\sigma$ in any bin. For Transit 4 the average change is approximately $1.5 \sigma$ per bin, but the reddest bin (which has the lowest S/N) changes at the $3 \sigma$ level. Due to this large inconsistency between systematics models, we exclude the reddest bin from our results.

Some authors use the residuals to their white light curve models to calibrate out common (i.e., wavelength-independent) systematics from each bin \citep[e.g.,][]{Sing13,Nikolov15,Huitson17}. In our case we find that such a correction is unwarranted. By eye, it is not apparent that our binned light curves share any common systematics (see Figures \ref{fig:130925}-\ref{fig:150927} in the Appendix). Furthermore, when we apply the common mode correction, we find that the impact on the resulting transit spectra is negligible compared to the uncertainties.

\subsection{Red noise} \label{sec:red_noise}

As discussed in Section \ref{sec:transit_fit}, we fit a parameter $\sigma_r$ to characterize the level of correlated (``red'') noise in our data. While the ratios in Table \ref{tab:transit_data} suggest a large amount of systematic noise, this does not necessarily imply \emph{correlated} noise; the systematic noise may still be captured by the white noise parameter.

\begin{figure}
    \centering
    \includegraphics[width=0.5\textwidth]{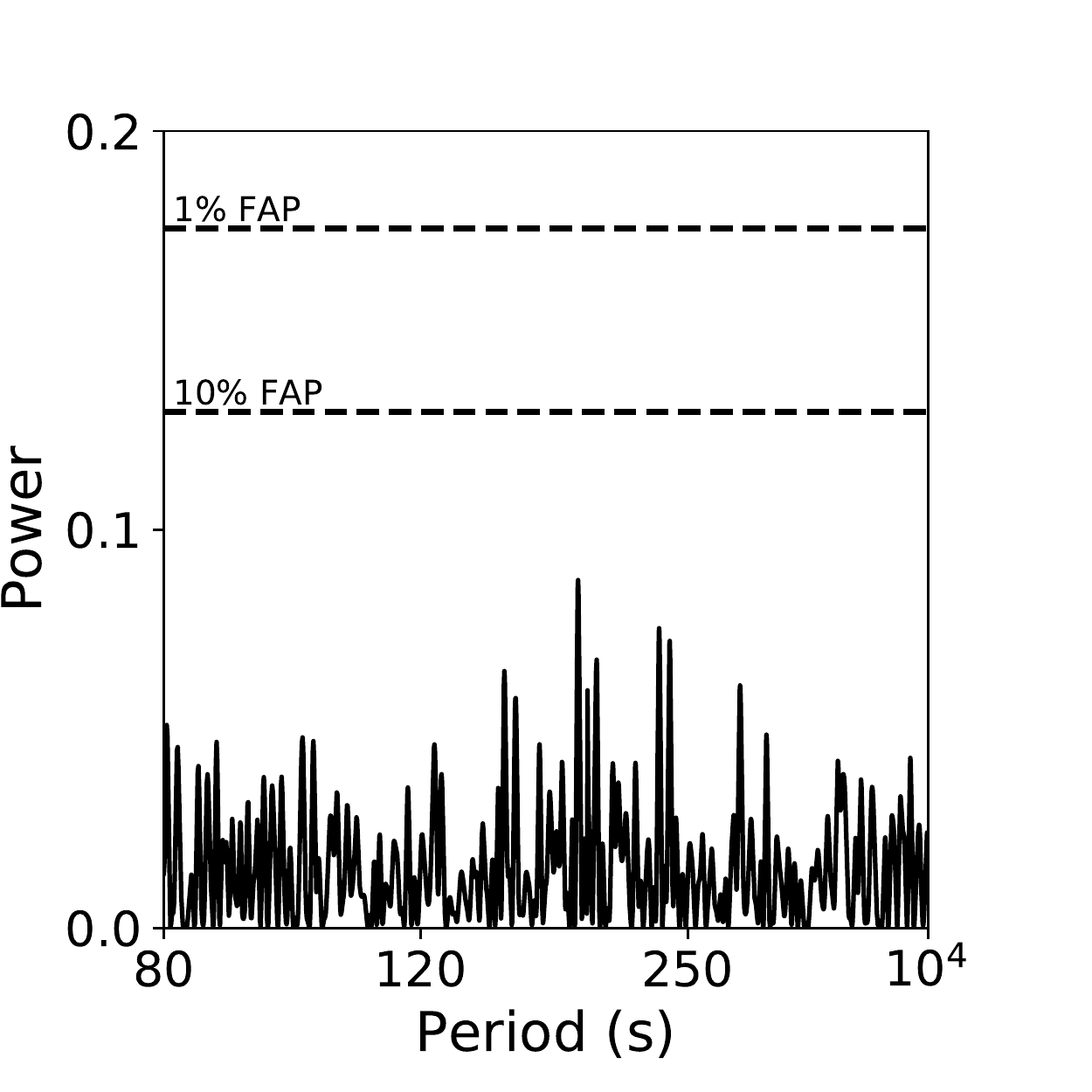}
   \caption{Lomb-Scargle periodogram of the residuals to the light curve model for the 610-630 nm bin of Transit 3. The lack of strong signals demonstrates that this light curve does not have strong periodic correlated noise across this range of frequencies.\label{fig:red_noise}}
\end{figure}

To assess the level of red noise in our data, we construct Lomb-Scargle periodograms of the residuals to each binned light curve model, then use bootstrap resampling to calculate the false alarm probability \citep[FAP; e.g.,][]{VanderPlas18} for signals with periods between the exposure times and the duration of our observations. Figure \ref{fig:red_noise} shows an example periodogram for one of the bins of Transit 3. We find that only 6 out of 82 binned light curves possess signals with FAP < 10\%, and only 2 have FAP < 1\%. This suggests that there are few strong periodic signals in our light curves. Nevertheless, we conservatively choose to apply the red noise parameterization, which results in a $\sim 15\%$ increase in our uncertainties.

\subsection{Effects of incomplete phase coverage}\label{sec:phase_coverage}
The light curves of Transits 1-3 all suffer from a partial lack of phase coverage, due either to poor weather, observing window constraints, or spot-crossing events which must be masked in the model fit. In principle, our MCMC approach to fitting the model should adequately incorporate this lack of information into the uncertainties on $R_p/R_s$. However, it is worthwhile to estimate the extent to which this missing information inflates our uncertainties.

The full light curve of Transit 4 allows us to investigate the effects of poor phase coverage in the previous three transits. Specifically, we remove data points from the binned light curves of Transit 4 to mimic the phase coverage of Transits 1-3, then fit for the transit depth and calculate the resulting increase in the uncertainties on the binned values of $R_p/R_s$. Our results are summarized as follows:

\begin{itemize}
    \item Mock Transit 1: We remove the pre-ingress baseline and most of the first half of the transit. This results in a 150\% increase in uncertainties, as well as a 4.5$\sigma$ decrease in the wavelength-averaged radius. The $\Delta R_p/R_s$ values change at the $>1 \sigma$ level, and the change appears to be wavelength dependent, introducing a slope on the order of $\sim 1 \sigma$ from blue to red.
    \item Mock Transit 2: We remove the points near mid-transit which correspond to a spot-crossing event. This results in a 50\% increase in uncertainties, and a $0.5 \sigma$ decrease in the average radius. The $\Delta R_p/R_s$ values change at the $\lesssim 0.2\sigma$ level.
    \item Mock Transit 3: First, we remove most of the post-egress baseline, which results in a 50\% increase in uncertainties with no considerable change in the average radius, and $\lesssim 0.5 \sigma$ changes in $\Delta R_p/R_s$. Next, we exclude only those points which correspond to the spot crossing event in Transit 3, finding effects of similar magnitude. Removing both portions of the light curve, however, does not increase the uncertainties any further.
\end{itemize}

The results for Transit 1 lead to concerns over whether the lack of a pre-ingress baseline could introduce a bias into the shape of our transit spectrum. Furthermore, the large uncertainties mean that this transit contributes little to the combined spectrum. We therefore opt to exclude the spectrum of Transit 1 from the combined result.

Meanwhile, the results for Transits 2 and 3 suggest that little bias has been introduced into the transit spectrum due to the lack of phase coverage in each, while the uncertainties on $\Delta R_p/R_s$ should be larger by $\sim 50\%$ than they would be given full phase coverage.

\subsection{Correcting for occulted spots} \label{sec:spot_correction}
The light curves of Transits 2 and 3 feature prominent spot occultation features which we exclude from the light curve model fit. While doing so allows us to fit the apparent transit depth without simultaneously modeling the spot, it does \emph{not} eliminate the impact of the occulted spot on the resulting transit spectrum. Since the spot is cooler than the rest of the photosphere, it breaks the transit model's assumption that the photosphere is azimuthally symmetric and amplifies the fitted transit depth in a wavelength-dependent manner \citep[e.g.,][]{Deming13}. In previous works we have referred to this as the transit light source effect (TLSE); here we derive an approximate correction (ignoring limb-darkening) for the TLSE due to a single spot.

Assuming the photosphere to have specific flux $F_{\lambda,\text{phot}}$ and radius $R_s$, the spot to have specific flux $F_{\lambda,\text{spot}}$ and radius $R_\text{spot}$, and the planet to have apparent radius $R_p$ and true radius $R_{p,0}$, then the observed relative transit depth will be
\begin{equation}
\left(\frac{R_p}{R_s}\right)^2 = \frac{F_{\lambda,\text{phot}}R_{p,0}^2}{F_{\lambda,\text{phot}}(R_s^2-R_\text{spot}^2)+F_{\lambda,\text{spot}}R_\text{spot}^2}
\end{equation} 
By re-arranging this equation, we derive a corrective factor for the transit depth\footnote{See also Equation 1 of \citet{McCullough14}, Equation 2 of \citet{Rackham18}.}:
\begin{equation}\label{eqtn:epsilon}
\epsilon_\lambda \equiv \left(\frac{R_p}{R_{p,0}}\right)^2 = \left[1- \left(1-\frac{F_{\lambda,\text{spot}}}{F_{\lambda,\text{phot}}}\right)\left(\frac{R_\text{spot}}{R_s}\right)^2 \right]^{-1}
\end{equation}

Later, in Section \ref{sec:starspots}, we find the feature in Transit 3 to be best described by a model of a single spot with radius $R_\text{spot}/R_s = 0.27_{-0.02}^{+0.03}$ and a temperature contrast of $T_s-T_\text{spot} = 100 \pm 5$ K. While this feature could in principle be due to a mixture of occulted spots and faculae, our data are insufficient to constrain more complex models, so we use the parameters of the single spot model for the purposes of correcting for the TLSE in this spectrum. We interpolate a PHOENIX \citep{Husser13} stellar photosphere model grid onto the values in Table \ref{tab:properties} to compute $F_{\lambda,\text{phot}}$ and $F_{\lambda,\text{spot}}$ at their respective temperatures, then calculate $\epsilon_\lambda$ from Equation \ref{eqtn:epsilon}.

\begin{figure}
\centering
\includegraphics[width=0.5\textwidth]{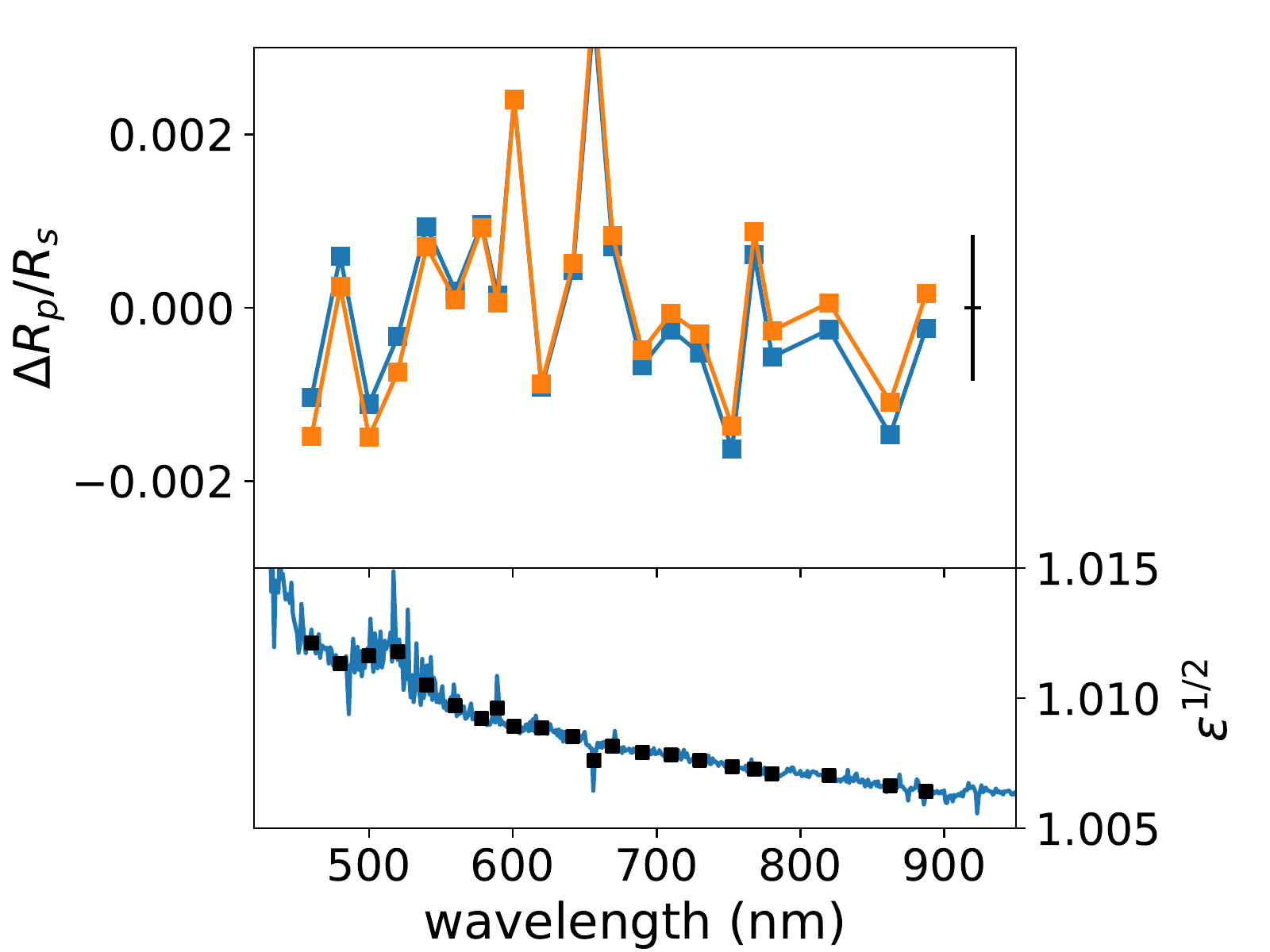}
\caption{(Top) The spectrum of Transit 3 before (blue) and after (orange) correcting for the large occulted spot. Both spectra have been mean-subtracted. The black error bar demonstrates the mean binned uncertainty, while the individual error bars are excluded for visibility. (Bottom) The radius contamination factor due to the spot, binned to 1 nm. The black squares are binned to match the spectrum above.\label{fig:correction}}
\end{figure}

In Figure \ref{fig:correction} we plot $\epsilon^{1/2}$, which is the corrective factor for the planet \emph{radius}, as well as the original and corrected spectra for Transit 3. The magnitude of the correction from 450 to 900 nm is approximately equal to the binned uncertainty. The corrective factor is listed in Table \ref{tab:data}, binned to match our final spectrum. This corrective factor is not applied to the spectrum of Transit 3 in Figure \ref{fig:specs_all}, but is applied before creating the combined spectrum in Figure \ref{fig:spec_combined}.

We do not derive a similar correction for Transit 2 because it is unclear (see Section \ref{sec:starspots}) whether this feature is best modeled by a small or large spot, and the slope introduced by the large spot model is six times steeper across the wavelength range than that of the small spot. Furthermore, the large spot model would only introduce a $\sim 0.3 \sigma$ increase in the measured radius from the reddest to bluest bins.

Finally, it is worth noting that this effect is just as prominent for spots outside of the transit chord. Equation \ref{eqtn:epsilon} can be further generalized by replacing $(R_\text{spot}/R_s)^2$ with $F_\text{het}$, the areal spot covering fraction of the photosphere, and using an average spot temperature to compute $F_{\lambda,\text{spot}}$.

\section{Starspots} \label{sec:starspots}
In Figure \ref{fig:lightcurves} we have identified spot-crossing features in the light curves of Transits 2 and 3, and we exclude these data points from our analysis of the transmission spectrum. Such features are common in transit light curves, and have previously been found in transit observations of WASP-4b \citep{Sanchis-Ojeda11}. The shape, central time, and magnitude of the feature carry information about the size, position, and contrast (or effective temperature) of the occulted spot, which are otherwise difficult to constrain. Transit spectroscopy allows us to measure the spot contrast at multiple wavelengths, permitting a more precise measurement of the spot's effective temperature than is available from wavelength-integrated light curves.

The measurement of spot properties is worthwhile for a variety of reasons. In the context of this paper, knowing the properties of an occulted spot allows us to correct for its effect on the spectrum of the transit during which it is observed. More generally, observations of multiple spots across an extended period of time permit a better understanding of the makeup of the photospheres of other stars.

\subsection{Spot modeling}
We use SPOTROD \citep{Beky14} to produce light curves of transits with one or two spot-crossing events. In addition to the usual transit parameters, SPOTROD models the position and radius (in units of stellar radii) as well as the average spot-to-stellar flux contrast for each spot. We employ PyMultiNest \citep{Buchner14} to sample the parameter space and to calculate the Bayesian log-evidences $\ln(Z)$ through which we can compare the different models, as discussed in Section \ref{sec:model_comparison}. This MultiNest implementation\footnote{\url{https://github.com/nespinoza/spotnest}} of SPOTROD has also been used to study spots observed during transits of WASP-19b \citep{Espinoza19}.

We use a second-order polynomial to detrend the white light curve from each night, then fit the combined spot and star model. We limit the uniform prior on spot size to $R_\text{spot}/R_s < 0.08$ or $0 < R_\text{spot}/R_s < 1$ to probe for small and large spots. Using the effective temperature in Table \ref{tab:properties} as a prior, the code then constructs and fits models for the flux contrast as a function of wavelength.

\begin{figure*}[ht]
\centering
\includegraphics[width=\wf\textwidth]{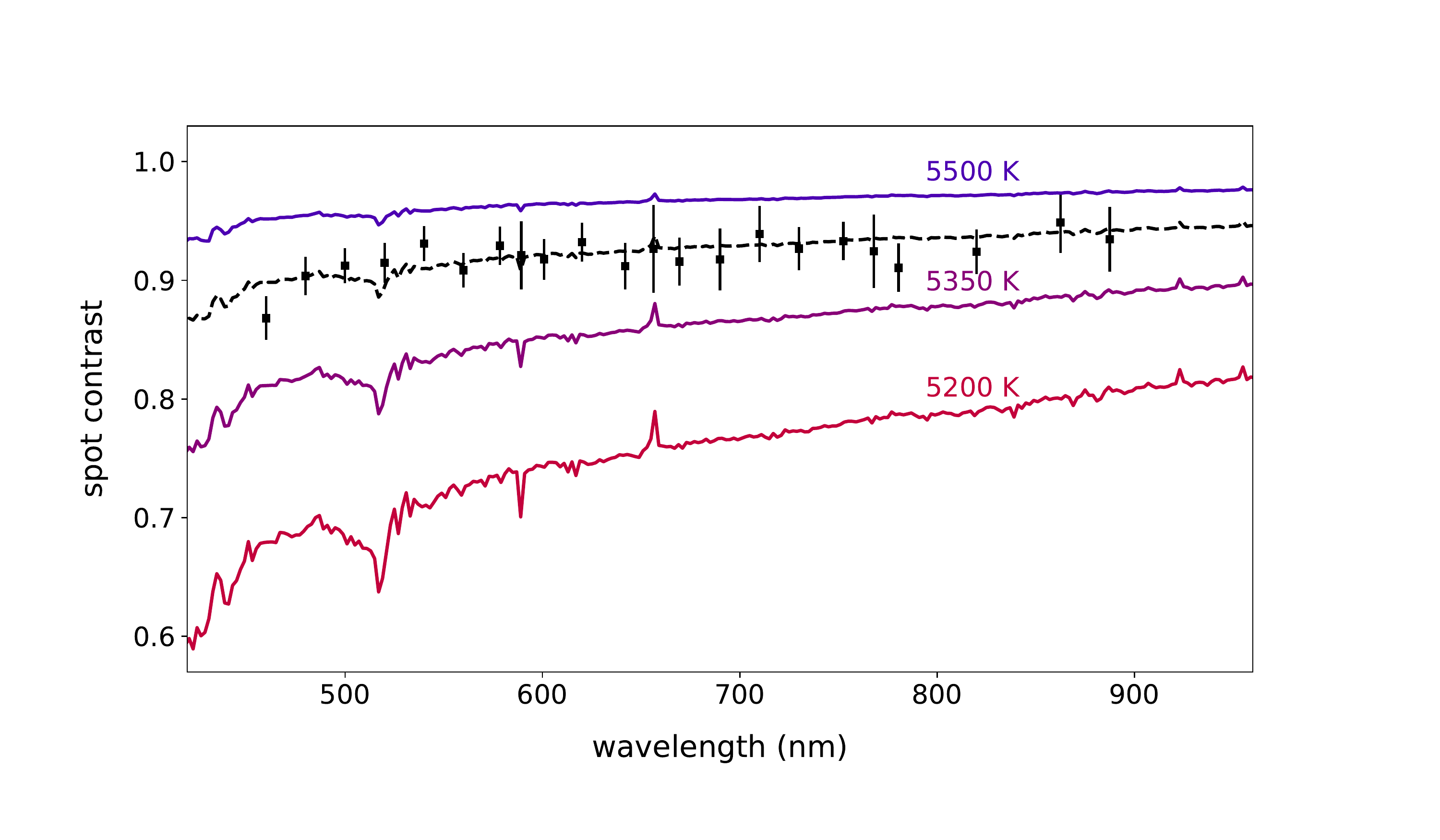}
\caption{Best-fit model for the occulted spot in the light curve of Transit 3. The data are best described by a single large, warm spot with a radius of $0.27\,R_s$ and a temperature contrast of 100 K.\label{fig:spot_model}}
\end{figure*}

The optimal model parameters for each spot feature are presented in Table \ref{tab:spots}. The precision of the Transit 2 data only permits us to fit the spot feature and spectrum in a single, white-light bin. For this night, we find that our data are consistent with one- and two-spot solutions, as well as large and small spots. We opt for the simpler single-spot models and present the fit for both a single large and single cool spot. The data for Transit 3 favor a single large spot versus multiple or smaller spots ($\Delta \ln(Z) > 2$). For this transit we are able to fit the spot spectrum to the same bins as our transmission spectrum, as shown in Figure \ref{fig:spot_model}.

The large spot model for Transit 3 seems inconsistent with smaller individual spots observed on the Sun; nevertheless, such wide features have been observed in transits before \citep[e.g.,][]{Espinoza19}. While our evidences do not favor a two-spot model, it is possible that the large, low-contrast spot is in fact a more complex active region consisting of multiple spots and/or faculae, as has been observed on the Sun. Regardless of the actual structure of the feature, its average contrast is well-described by a single spot with $T_s-T_\text{spot} = 100$ K, and we choose this model to calculate the correction in Section \ref{sec:spot_correction}.

\begin{deluxetable}{lllll}
\tablecaption{Best-fit parameters for the spot-crossing events in the light curves of Transits 2 and 3. The data for Transit 2 are consistent with both a small (2a) and large spot (2b). Median and $1\sigma$ confidence intervals are reported. \label{tab:spots}}
\tablehead{& $R_\text{spot}/R_s$ & $f_\text{spot}/f_s$ & $T_s-T_\text{spot}$ (K) & $T_\text{spot}$ (K)}
\startdata
Transit 2a & $0.05^{+0.01}_{-0.01}$ & $0.49^{+0.20}_{-0.28}$ & $880 \pm 360$ & $4670 \pm 360$ \\
Transit 2b & $0.15^{+0.11}_{-0.07}$ & $0.79^{+0.11}_{-0.18}$ & $410 \pm 280$ & $5140 \pm 280$ \\
Transit 3  & $0.27^{+0.03}_{-0.02}$ & $0.91^{+0.01}_{-0.01}$ & $100 \pm 5$ & $5442 \pm 50$ \\
\enddata
\end{deluxetable}

\subsection{Spot sizes and temperatures} \label{sec:spot_sizes}
Later, in Section \ref{sec:atmosphere}, we assume a spot temperature in order to model the effects of unocculted spots on the transit spectrum. The temperature we assume reflects small, cool spots resembling those on the Sun, but the results for Transit 3 in Table \ref{tab:spots} appear to suggest the existence of large, warm features on the photosphere of WASP-4. In short, we do not believe that the spots discovered in our transit light curves are necessarily representative of the spots present elsewhere in the photosphere or below our detection threshold. We offer these arguments as justification:

\begin{enumerate}
\item \cite{Sanchis-Ojeda11} find evidence for a persistent small spot in multiple transit observations of WASP-4b. In this case, they find a spot with radius $R_\text{spot}/R_s \approx 0.05$ and temperature contrast $T_s-T_\text{spot} \approx 600$ K, suggesting that more Sun-like spots do exist on WASP-4.

\item Cool spots have previously been discovered through Doppler imaging of active G and K dwarfs \citep[e.g.,][]{ONeal98,ONeal04,Strassmeier09}. However, this method of spot detection is subject to biases regarding spot size and temperature.

\item The feature occulted during Transit 3 may indeed have a more complex structure consisting of cool spots and hot faculae which the precision of our data does not allow us to resolve.

\item While the precise radius and temperature of the spot observed during Transit 2 are largely unconstrained, we can say with at least $1\sigma$ confidence that the spot is smaller and cooler than the feature observed during Transit 3. Therefore, even our data suggest that small spots may be present in the photosphere of WASP-4.

\end{enumerate}

It is nevertheless possible that the distribution of spot sizes on WASP-4 favors larger and hot spots, in which case the spectral contamination signal would be less severe. However, our use of the contamination model is not meant to accurately account for the actual surface heterogeneity, but rather to demonstrate that this aspect should not be ignored in determining the planet's transmission spectrum.

\section{Atmospheric retrieval} \label{sec:atmosphere}

To interpret the combined transmission spectrum in Figure \ref{fig:spec_combined}, we employ a Bayesian atmospheric retrieval code based on PyMultiNest which has previously been used to study WASP-19b \citep{Espinoza19}. Following \cite{Betremieux17} and \cite{Heng17}, our atmosphere model includes two regions: an optically thick base region (which could also be interpreted as a cloud layer) with radius $(R_p/R_s)_0$ and reference pressure $P_0$, and an isothermal optically thin region above with temperature $T$. The components of the optically thin region may include either or both of (i) a set of atomic and molecular species and (ii) a scattering haze defined by a cross-section power law $\sigma = a\sigma_0 \left(\lambda/\lambda_0\right)^\gamma$, where $\sigma_0 = 5.31\times10^{-27}$ cm$^2$ is the Rayleigh scattering cross-section of H$_2$ at the reference wavelength $\lambda_0 = 350$ nm \citep{MacDonald17}, and $a$ is a dimensionless enhancement factor. We constrain $\gamma$ to be between 0 (uniform opacity) and $-4$ (Rayleigh scattering), which should span the range of physical scattering processes.

As we have previously argued in \cite{Rackham17,Rackham18}, it is important that analyses of transmission spectra account for the transit light source effect (TLSE) which can be introduced by heterogeneous features (spots and faculae) on the stellar surface. We incorporate a three-parameter model for the stellar photosphere into our retrieval code, fitting this model simultaneously with that of the planet's atmosphere. Section \ref{sec:heterogeneity} discusses this model in more detail.

The parameters and priors for each component of the model are listed in Table \ref{tab:priors}. We consider eight models for the atmosphere, including all possible combinations of a cloud deck, a scattering haze, and Na and/or K absorption signals. Each of these is fitted independently or alongside a photosphere model (+1 free parameter, $F_\text{het}$) for a total of sixteen models.

\subsection{Model comparison} \label{sec:model_comparison}
To compare between our models, we compute their relative Bayes factors. A thorough introduction to the use of Bayes factors in astronomy is given by \cite{Trotta08}, and their use in transit spectroscopy is well-established \citep[e.g.,][]{Benneke13}. Here, we provide a brief overview.

Given a model $M$ with parameters $\theta$, the optimal values of $\theta$ to describe data values $x$ are those which maximize Bayes' equation:
\[
P(\theta | x,M) = \frac{P(x | \theta,M) P(\theta|M)}{P(x|M)}
\]
where $P(x | \theta,M)$ is the likelihood function and $P(\theta|M)$ is the prior distribution of $\theta$. Most Bayesian model fitting algorithms seek to maximize the likelihood function, but the maximum likelihoods of two different models cannot be directly compared.

To compare two models, we first compute their Bayesian evidences, defined as:
\[
Z_M \equiv \int P(x|\theta,M) P(\theta|M) d\theta
\]
The evidence for a model is high if the data are well-described by a large region of the prior parameter space. The Bayes factor between two models is defined as the ratio of their evidences:

\[
\ln(B) \equiv \ln(Z_2)-\ln(Z_1)
\]

In general, model 2 is weakly favored over model 1 if $0 < \ln(B) < 2$, and strongly favored if $\ln(B) > 5$. If $\ln(B)<0$ then model 2 is disfavored to model 1. Complex models will be favored over simpler models only if the evidence justifies the additional parameters. However, a preference towards simpler models does not necessarily mean that the complex models are incorrect, but rather that they are not justified by the data at hand.

Our retrieval algorithm computes the Bayesian evidence parameter for each of the fitted models. In the following sections, we cite the Bayes factor $\ln(B)$ for each model with respect to the model for a uniform opacity atmosphere with no contamination from the photosphere (i.e., a flat line). In general, we find that the more complex models have $\ln(B)<0$, meaning that they are \emph{dis}favored versus a flat line.

\begin{deluxetable*}{l|llll}[ht]
\tablecaption{Parameters for our combined photosphere and atmosphere models. Not all parameters are incorporated in all models; rather, Bayesian log-evidences are used to compare separate models.\label{tab:priors}}
\tablehead{\colhead{Model component} & \colhead{Parameter} & \colhead{Units} & \colhead{Description} & \colhead{Prior distribution}}
\startdata
Base		 		& $(R_p/R_s)_0$ 	& -  	& Radius corresponding to the top of the cloud layer or $\tau \gg 1$ 	& Uniform(0.1,0.2)\\
					& $P_0$\tdg 		& bar 	& Reference pressure at $(R_p/R_s)_0$ 									& Log-uniform($10^{-6}$,$10^6$) \\
               	 	& $T$\tdg			& K		& Average temperature of the optically-thin region						& Uniform(1000,2400) \\ \hline
Atomic features 	& $X$ 				& - 	& Mixing ratio of species X 											& Log-uniform($10^{-30}$,1)\\ \hline
Haze				& $a$				& - 	& Amplitude of the haze cross-section power law								& Log-uniform($10^{-20}$,$10^{10}$) \\
					& $\gamma$			& -		& Index of the haze cross-section power law								& Uniform(-4,0) \\ \hline
Stellar photosphere & $T_\text{occ}$ 	& K 	& Average temperature of the transit chord\tablenotemark{$\ddagger$}	& Fixed(5540) \\ 
					& $T_\text{het}$ 	& K 	& Average temperature of the heterogeneous surface features 			& Fixed(4200) \\
                    & $F_\text{het}$ 	& - 	& Fraction of the unocculted photosphere covered by spots		 	& Uniform(0,1) \\ 
\enddata
\tablenotetext{\dagger}{The temperature and reference pressure are not modeled in the case of a uniform opacity atmosphere.}
\tablenotetext{\ddagger}{This excludes any occulted features which can be identified in the light curve.}
\end{deluxetable*}

\begin{figure*}[p]
\centering
\includegraphics[width=\textwidth]{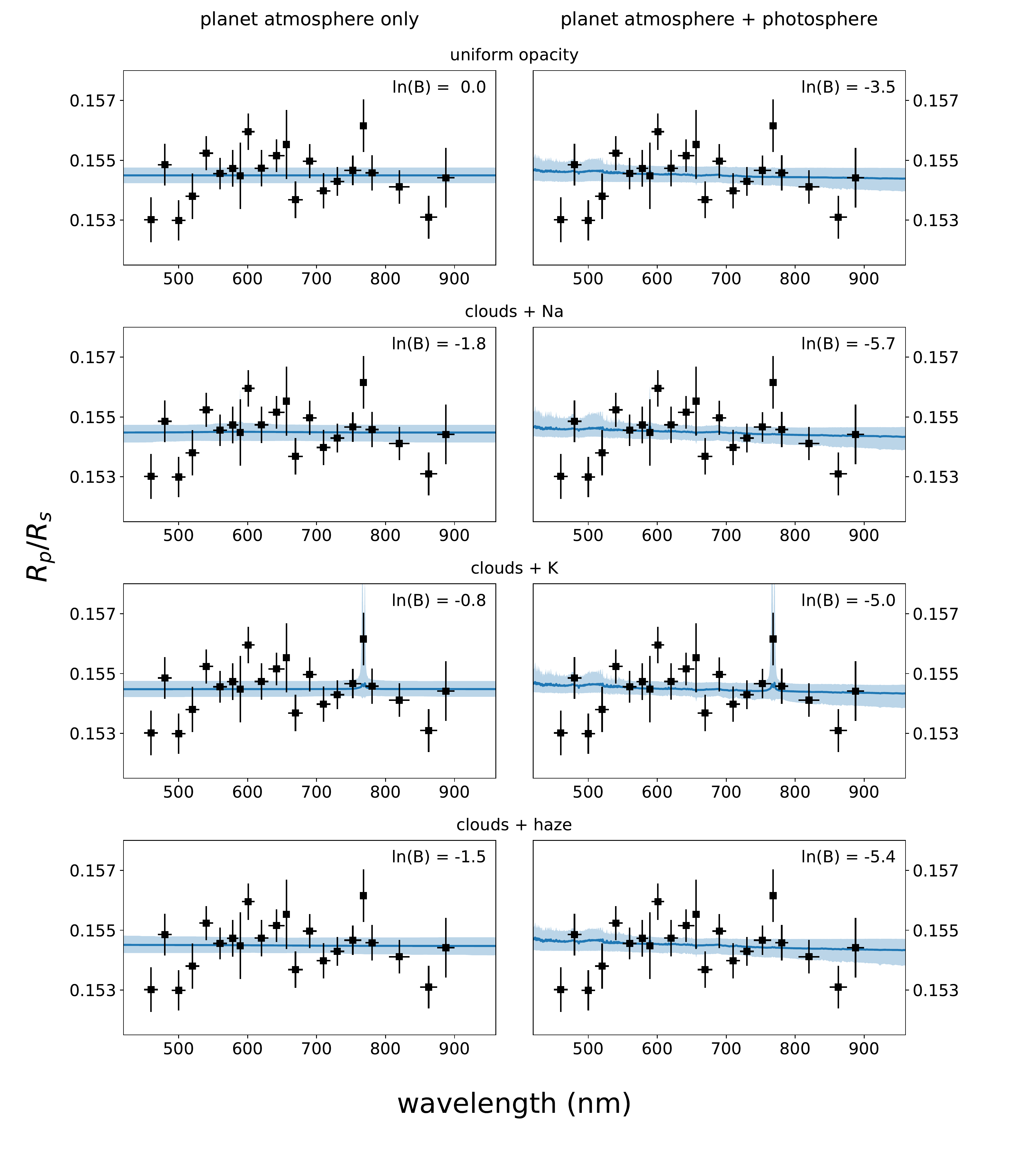}
\caption{A subset of fitted models for the observed transmission spectrum of WASP-4b; four separate models for the planetary atmosphere are presented alone (left) or combined with a model for the heterogeneous stellar photosphere (right). The atmosphere models, described in more detail in Section \ref{sec:atmosphere}, include (top) a uniform opacity atmosphere, (middle) a cloud deck and $Na$ absorption feature, a cloud deck and $K$ absorption feature, and (bottom) a cloud deck and scattering haze. The shaded region represents the 95\% confidence interval around the mean model fit. Bayes factors are given for each model relative to the flat line model (upper left). A constant offset of +0.1545 has been added to the combined spectrum. \label{fig:retrieval}}
\end{figure*}

\subsection{Stellar heterogeneity} \label{sec:heterogeneity}

\begin{figure*}[ht]
\centering
\includegraphics[width=\textwidth]{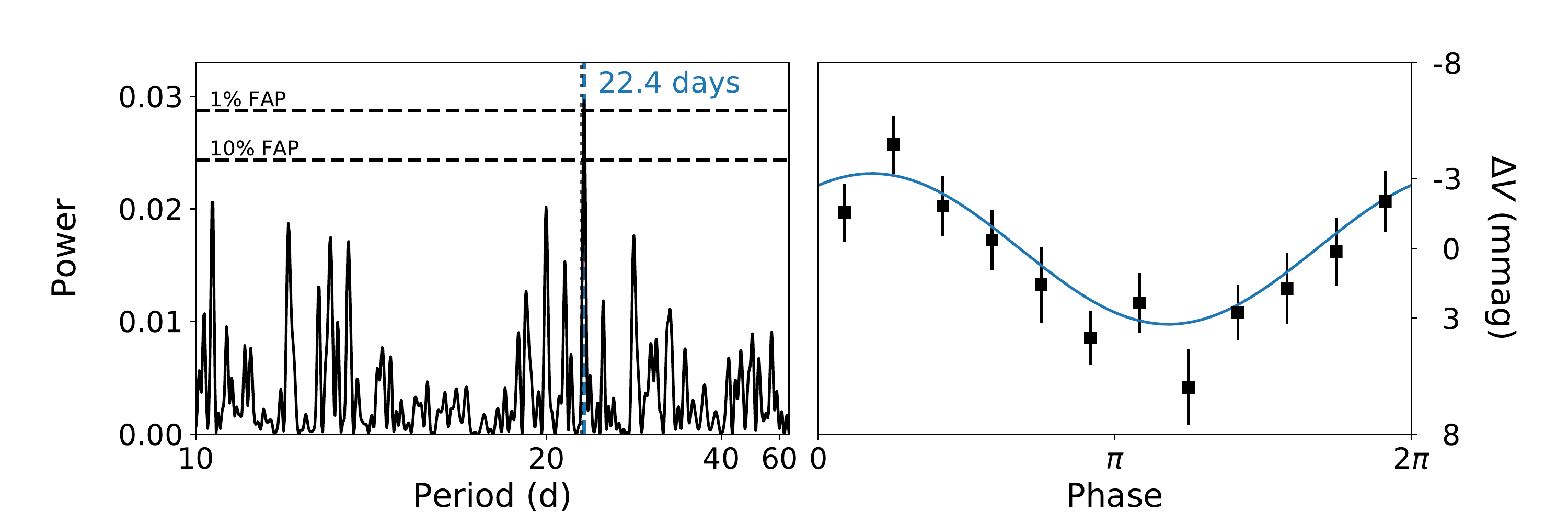}
\caption{(Left) Lomb-Scargle periodogram for the $V$-band magnitude of WASP-4 from ASAS-SN photometry. The largest peak, marked with a blue line, corresponds to the previously measured rotation period, marked with a black dotted line. The horizontal lines denote the false alarm probability calculated through the bootstrapping method. (Right) Fitted model for the phase-folded light curve, assuming a period of 22.4 days. The data are binned for visibility. \label{fig:periodogram}}
\end{figure*}

Assessing the level of TLSE contamination is critical to correctly interpreting high-precision transmission spectra of transiting exoplanets \citep[][]{Apai18}. This is particularly important for investigating features that could have a planetary or stellar origin, such the He 10,833~\AA\ line \citep{Spake18} or the Na and K alkali lines \citep{Cauley18}. We have previously demonstrated this effect in the spectrum of GJ~1214b \citep{Rackham17}, which has a featureless near-IR spectrum but a visible spectrum that is apparently offset below the near-IR transit depths. In this study, we found that for reasonable assumptions about the star's activity, the planetary transmission spectrum of GJ~1214b can be shown to be flat in the visible as well. We also detect the signature of stellar activity in one of our \imacs\ transit data sets for WASP-19b \citep{Espinoza19}.

WASP-4 is of slightly later spectral type than the Sun, and multiple spot-crossing events have been observed during previous transit observations of WASP-4b \citep{Sanchis-Ojeda11} as well as in two of our data sets. To assess the photometric variability, we retrieve four years of photometric monitoring data from the ASAS-SN online database\footnote{\url{https://asas-sn.osu.edu/}} \citep{Shappee14,Kochanek17}. We remove outliers with a 5$\sigma$ cut and create a Lomb-Scargle periodogram to search for periodic signals between 1 and 100 days, calculating the false alarm probability as in Section \ref{sec:red_noise}. As shown in Figure \ref{fig:periodogram}, we find the largest peak at 22.4 days, and while the significance of the detection is only moderate, it corresponds well with the rotational period of 22.2 days which \cite{Sanchis-Ojeda11} measure from consecutive spot-crossing events. This is to be expected if the photometric variability is dominated by the rotation of spots and faculae in and out of view. Modeling the variability as a sine curve and assuming a period of 22.4 days, we perform a least-squares fit to the phase folded data to find a peak-to-trough variability amplitude of $\sim 6$ mmag (0.6\%). These results suggest that the surface of WASP-4 is moderately heterogeneous, so any analysis of the transmission spectrum of WASP-4b which ignores stellar contamination may be overly optimistic. Furthermore, since the spot covering fraction can not be determined from the variability alone \citep{Rackham18}, the photosphere heterogeneity should be modeled alongside the atmosphere.

In the retrieval code described above, we characterize the heterogeneity with three parameters: $T_{occ}$, the effective temperature of the transit chord, $T_{het}$, the mean effective temperature of the heterogeneous features, and $F_{het}$, the fraction of the unocculted photosphere which is covered in spots. We then compute the effect on the observed transit depth following Section \ref{sec:spot_correction} and Equation \ref{eqtn:epsilon}, but replacing the area ratio $(R_\text{spot}/R_s)^2$ with a covering fraction $F_{het}$.

We test more complex parameterizations which include multiple types of unocculted heterogeneities \citep[e.g.,][]{Zhang18}, but determine from the evidences that they are not warranted by the data.

Given the quality of our data, we find it necessary to fix some parameters of the heterogeneity model to reasonable values. In Table \ref{tab:priors} we fix the spot temperature to match typical spots on the Sun \citep[e.g.,][and references therein]{Solanki03}, and the transit chord temperature to the measured effective temperature of the photosphere.

$F_\text{het}$ is allowed to vary over all possible values, and serves as a simple measure of the level of stellar contamination in the transmission spectrum. However, we note that $F_\text{het}$ represents an \emph{enhancement} over the effect of the large occulted spot in Transit 3, which has already been corrected for in Section \ref{sec:spot_correction}.

\subsection{Retrieval results}
In Figure \ref{fig:retrieval} we present four of our atmosphere models with and without photosphere contamination. In Table \ref{tab:fits} we list the Bayes factors for all sixteen of the models tested.

\subsubsection{Favored models}
The evidences favor a uniform opacity model for the atmosphere, with other atmosphere models being disfavored by $\ln(B) = -1$ to -2. Even when the haze is included, the amplitude $a$ of the haze opacity is small; the posterior distributions do not converge well enough from the log-uniform prior to offer a meaningful upper limit, but the mode of the distribution for $a$ lies between $10^{-18}$ and $10^{-12}$, which is lower than in the prior distribution.

Atmosphere models including Na and/or K are disfavored versus uniform opacity models by $\ln(B) = -1$ to -2.  Even though the narrow K bin is elevated above the continuum by $\sim 1\sigma$, ultimately the presence of potassium is not warranted by the data. We note that we cannot place upper limits on the abundances of Na or K, as these are degenerate with the reference pressure.

The models which include a heterogeneous photosphere are strongly disfavored by $\ln(B) = -5$ to -6 versus their counterparts with a homogeneous photosphere. While it is quite likely that $F_\text{het} > 0$, the spectrum \emph{alone} does not reveal evidence for it beyond what we have already corrected for in Section \ref{sec:spot_correction}. The effect of even a low spot covering fraction ($< 3\%$) is apparent in Figure \ref{fig:retrieval}, and is degenerate with the effect of a scattering haze. For this reason, we recommend the joint modeling of the photosphere and atmosphere in future analyses of optical transmission spectra, and the use of Bayesian evidences or likelihood parameters for model selection \citep[see also][]{Pinhas18}.

\subsubsection{Constraints on spot covering fraction} \label{sec:covering_fraction}
The parameters in the clear and hazy atmosphere models do not converge enough from their priors to offer meaningful constraints. However, in the special case of a uniform opacity atmosphere with stellar contamination, we can place a 95\% upper limit on the spot covering fraction, $F_\text{het} < 3.4\%$. We caution that this is under the assumption of 4200 K spots and does not factor in the likely diversity of spot and faculae temperatures on the stellar surface. Instead, it should be taken as a limit on the net spectral effect of contamination.

\begin{deluxetable}{lccr}
\tablecaption{Bayes factors for the full suite of atmosphere models, some of which are shown in Figure \ref{fig:retrieval}. The two columns represent atmosphere-only (A) and combined atmosphere-photosphere (A+P) models. $\ln(B)$ is calculated relative to the flat line model; since $\ln(B) < 0$ for the more complex models, they are all disfavored versus a flat line. \label{tab:fits}}
\tablehead{\colhead{Model} & \colhead{$\ln(B)_A$} & \colhead{$\ln(B)_{A+P}$} &}
\startdata
uniform opacity & 0.0 & -3.5 \\
clouds + Na & -1.8 & -5.7 \\
clouds + K & -0.8 & -5.0 \\
clouds + Na,K & -1.0 & -5.4 \\
clouds + haze & -1.5 & -5.4 \\
clouds + haze + Na & -1.9 & -6.2 \\
clouds + haze + K & -1.9 & -5.4 \\
clouds + haze + Na,K & -1.9 & -6.2 \\
\enddata
\end{deluxetable}

\subsection{Correcting vs. fitting the contamination in Transit 3}
In Section \ref{sec:spot_correction} we detail a method for correcting the contamination due to the cool feature which was occulted by the planet during Transit 3. In this retrieval, however, we also include a varying parameter $F_{het}$ and fixed parameter $T_{het}$ to characterize the heterogeneity of the photosphere. We offer the following argument to justify our decision to visit the heterogeneity twice.

First, while the detailed structure of the occulted feature from Transit 3 is unclear, it is well-described by a circular spot model with a constant temperature, and this is the same model which we use to correct for its contamination. By excluding this correction, we would not be leveraging all of the available information from our light curve. However, $F_{het}$ must still be non-zero to describe the remaining unocculted features.

Second, we demonstrate that this correction is accounted for during the atmospheric retrieval. As a test, we fit the model of a uniform opacity atmosphere with a heterogeneous photosphere (Figure \ref{fig:retrieval}, top-right), this time fixing $T_{het} = 5442$ K to match the temperature of the occulted feature. When we fit the uncorrected combined spectrum, we find that the median of the posterior distribution for $F_\text{het}$ is larger than for the corrected combined spectrum. The difference is consistent with a single spot of size $R_{spot}/R_s \approx 0.22$. Similarly, when the retrieval is applied to the corrected and uncorrected spectra from Transit 3 only, the difference in the median value of $F_{het}$ is consistent with a single spot of size $R_{spot}/R_s \approx 0.33$. Both of these are similar to the spot size assumed in the correction, $R_{spot}/R_s = 0.27^{+0.03}_{-0.02}$. This indicates that the effect of the correction is carried forward into the retrieval; we are not over-correcting for the effect of the spot.

\section{Comparison to published results}
\subsection{\gmos}
\cite{Huitson17} (hereafter H17) have previously published an optical transmission spectrum of WASP-4b from four combined transit observations with \gmos, and find that their data favor a uniform opacity model. In this section, we assess the agreement of their results with our data from IMACS. In Figure \ref{fig:gmos} we compare the red and blue GMOS spectra with the combined spectrum from this work, and find that the slopes of our spectra are generally in agreement.  

\subsubsection{Absorption features}

The H17 data reveal tentative evidence for Na absorption at 589 nm. Our data reveal no evidence for such a feature, but we concede that the quality of our data in the blue is not sufficient to definitively rule out an atomic feature of small optical depth.

H17 exclude bins in their spectra which are coincident with terrestrial telluric $O_2$ absorption, finding that their correction for differential atmospheric refraction was not reliable in this wavelength range. Magellan is equipped with an ADC so our data do not require this correction (see Section \ref{sec:stability}). As a result, we are able to test for a K I feature in the same region, and find little evidence as discussed in Section \ref{sec:atmosphere}.

\subsection{\imacs}
\begin{figure*}[p]
\centering
\includegraphics[width=\wf\textwidth]{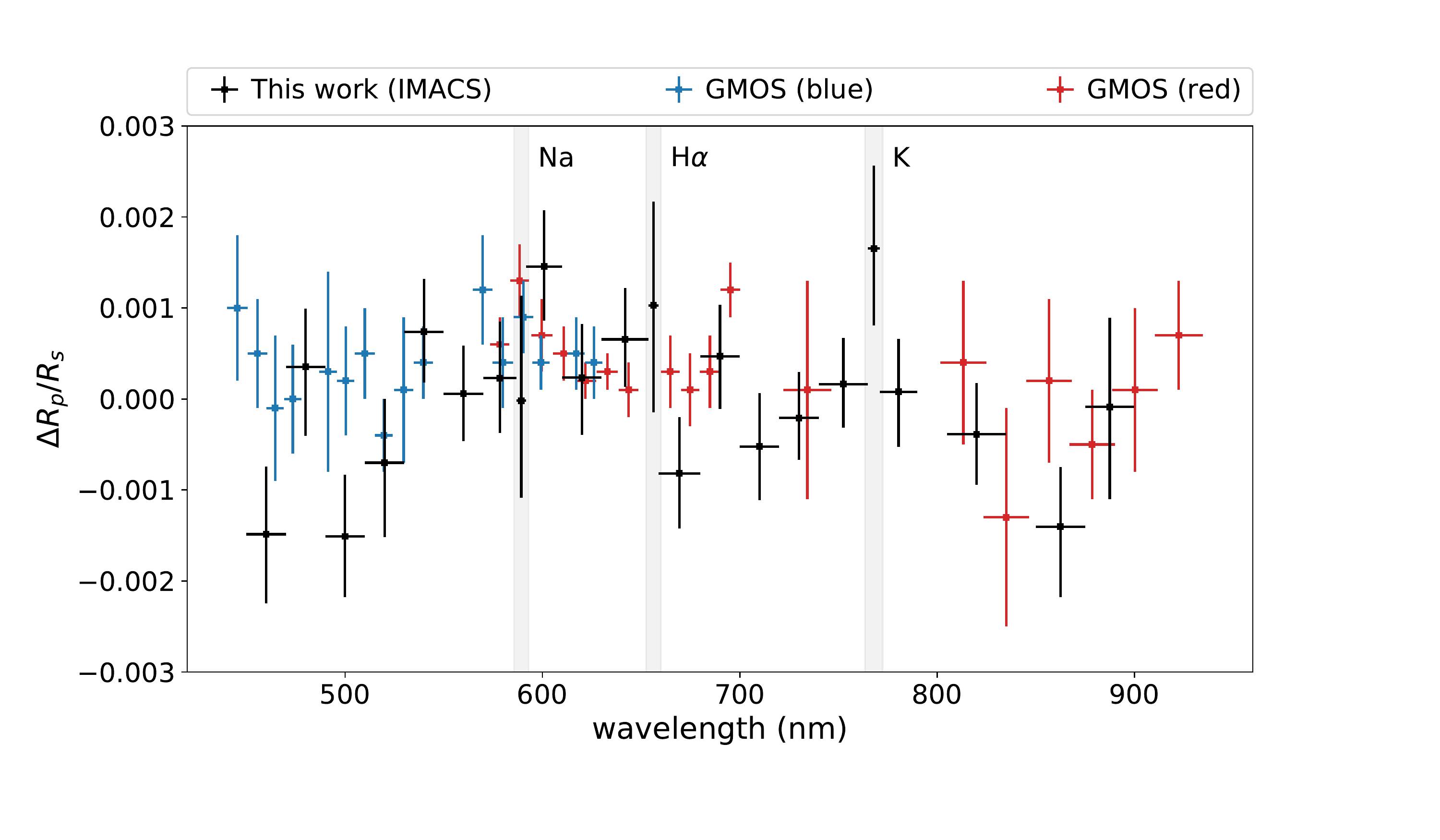}
\includegraphics[width=\wf\textwidth]{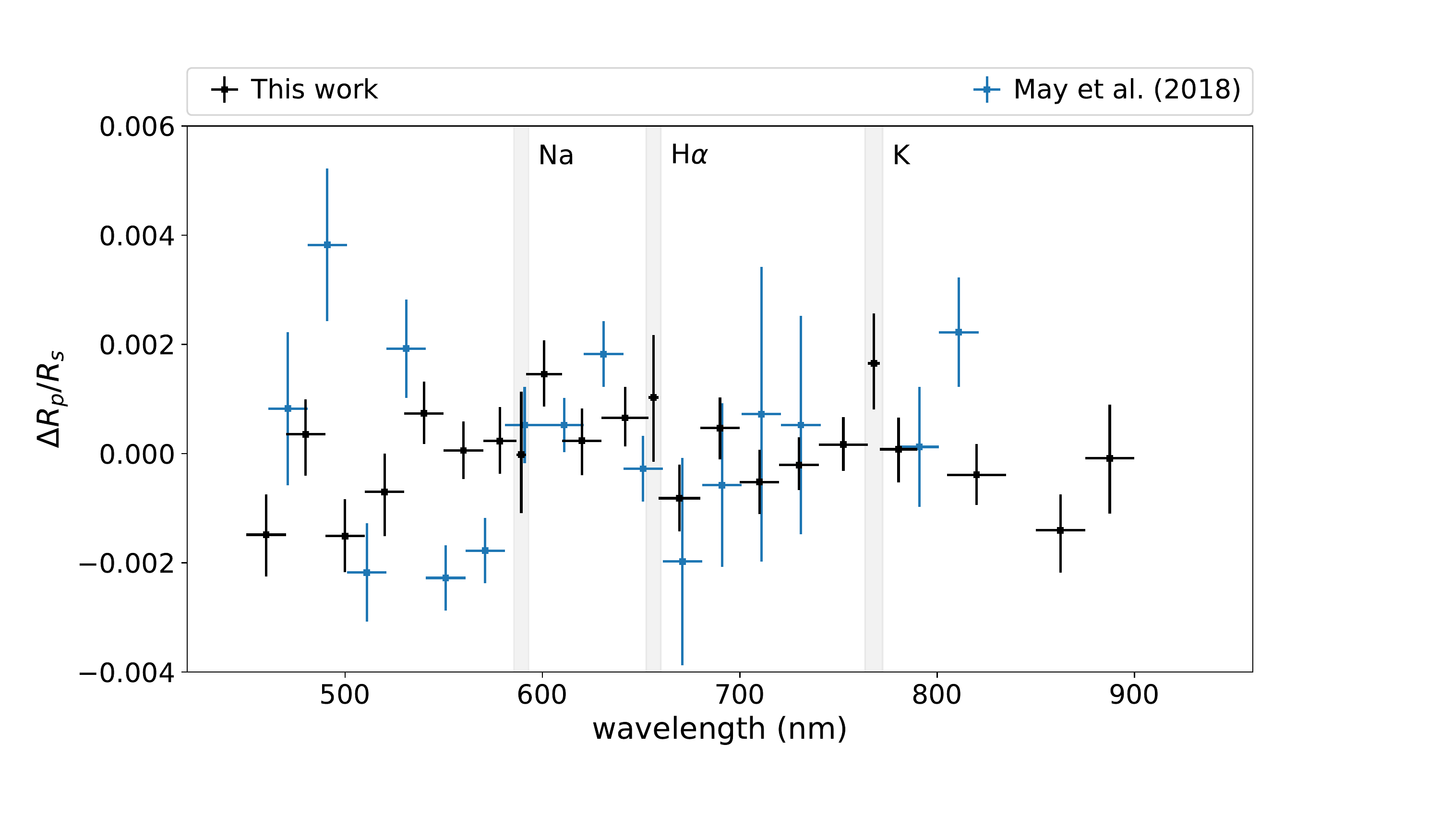}
\caption{A comparison of our combined transmission spectrum to the spectrum of (top) \cite{Huitson17} (Table 4) as observed with \gmos\ and (bottom) \cite{May18} (Table 5) as observed with IMACS. The wavelengths of potential atomic features are highlighted. The weighted mean is subtracted from each of the three data sets.\label{fig:gmos}}
\end{figure*}

\cite{May18} (hereafter M18) used \imacs\ to study this target with the same grism as we used for Transits 1-3, and they also find no evidence for features in the transit spectrum. Figure \ref{fig:gmos} demonstrates that our results are in agreement.

\subsubsection{Lack of a spot crossing event}
The transit observed by M18 occurred on 19 August 2015, approximately 5.4 days after our Transit 3, though their light curves show no evidence for a large occulted photosphere feature. This is not unexpected: the planet's period is prograde and nearly equatorial, and the stellar rotation period is $\sim 22.2$ days \citep{Triaud10, Sanchis-Ojeda11}. This implies that the feature observed during Transit 3, which was occulted in the second half of the transit, would have rotated $\sim 90$ degrees and off the projected stellar disk within the period of time between the transits.

\subsection{Combined analysis}
We combine the data from H17 and M18 with our own by subtracting the weighted mean from each spectrum, then repeat the retrieval analysis of Section \ref{sec:atmosphere}. Table \ref{tab:fits_combined} displays the Bayes factor relative to a flat line for each for the 2x8 models. As before, we find that most atmosphere models are disfavored versus a uniform opacity model by $\ln(B) = -1$ to $-2$, but models including Na are slightly less disfavored than in Table \ref{tab:fits}. This indicates that there is \emph{more} evidence for Na absorption in the combined data set than in ours alone, but still not enough to justify its inclusion.

Models which include the photosphere are disfavored by $\ln(B) \sim -5.5$. In the case of a uniform opacity atmosphere, the 95\% upper limit on the spot covering fraction is $F_\text{het} < 1.8\%$, which is smaller than the value we report in Section \ref{sec:covering_fraction}. However, we caution that the red and blue GMOS spectra were observed during separate transits, which may mask the slope introduced by TLSE contamination.

\begin{deluxetable}{lccr}
\tablecaption{Bayes factors for the full suite of atmosphere models (similar to Table \ref{tab:fits}) where we have included data from H17 and M18. \label{tab:fits_combined}}
\tablehead{\colhead{Model} & \colhead{$\ln(B)_A$} & \colhead{$\ln(B)_{A+P}$} &}
\startdata
uniform opacity & 0.0 & -3.8 \\
clouds + Na & -0.2 & -5.5 \\
clouds + K & -1.3 & -5.7 \\
clouds + Na,K & -0.7 & -4.9 \\
clouds + haze & -1.7 & -6.4 \\
clouds + haze + Na & -1.2 & -6.0 \\
clouds + haze + K & -1.6 & -6.3 \\
clouds + haze + Na,K & -1.6 & -5.9 \\
\enddata
\end{deluxetable}

\section{Featureless Atmosphere}

Our transit spectra, in combination with the spectra by H17 and M18, suggest that WASP-4b displays a mostly featureless (uniform opacity) spectrum without a strong optical spectral slope or alkaline absorption (Na~{\sc I} or K~{\sc I} lines). Although more observations will be required to place tighter constraints on the optical spectrum to verify this, in the following we will explore what a uniform opacity spectrum would suggest for WASP-4b. 

Cloud-free, broadly solar-composition atmospheres are predicted to display strong absorption at visible wavelengths by alkali metal atoms (Na~{\sc I} or K~{\sc I} doublets, \citealt[][]{Seager00b,Sudarsky00,Seager00a}). These prominent (deep and broad) features have been observed in the spectra of multiple transiting hot Jupiters \citep[e.g.,][]{Charbonneau02}. The depths of the truncated alkali features are often used as proxy for the atmospheric pressures probed \citep[e.g.,][]{Sing15}. The emerging evidence argues that transit sightlines are often limited by cloud decks, which then lead to truncated alkali line cores or, for very low-pressure particles, even the absence of detectable alkali absorption. 

Another marked deviation from a flat (featureless, zero-slope) visible spectrum would arise in a clear atmosphere from Rayleigh scattering of the starlight by molecules or very small particles. Rayleigh scattering is more efficient at smaller wavelengths, resulting in apparently larger planet diameters (i.e., a lower pressure level $\tau=1$ surface) at shorter wavelengths \citep[e.g.,][]{Pont08}. The actual wavelength-dependence of the Rayleigh-scattering in a given transiting exoplanet atmosphere will depend on the particle size distribution in its upper atmosphere and can vary by orders of magnitude (e.g., \citealt[][]{Heng17}).
However, visual-wavelength slopes in transiting exoplanets may also be introduced by the transit light source effect described in Section \ref{sec:spot_correction}; given this consideration, the simultaneous presence of visual slopes and alkali line absorption are considered to be the most robust indicators of clear (particle-free) upper atmospheres.

Therefore, with multiple transit spectra suggesting the { lack of} alkali absorption {\em and} the lack of a strong visual slope for WASP-4b, it is important to consider the possibility that WASP-4b's upper atmosphere is not clear but contains particles at high-altitude. Given this strong possibility, in the following we explore the possible nature of these particles and the mechanisms that may inject particles into the upper atmosphere. 
Based on \emph{Spitzer}/IRAC eclipse (dayside emission) measurements in the [3.6] and [4.6] filters, \citet[][]{Beerer11} found that the best-fit blackbody temperature of WASP-4b's dayside is 1,700~K. This temperature is close to the radiative equilibrium estimate of 1,650~K (assuming zero albedo) and significantly lower than 2,000~K, the temperature that would correspond to a zero-albedo, dayside-emission-only case. 

The dayside temperatures of WASP-4b are hot enough to vaporize less refractory grains and -- in a smaller fraction of the hemisphere centered on the sub-stellar point -- it is likely hot enough to vaporize metal-oxide and silicate grains. Cloud formation through the evaporation and condensation of refractory grains is common and relatively well-studied in non-irradiated brown dwarfs \citep[e.g.,][]{Apai2013,Buenzli14,Metchev2015} and directly imaged exoplanets \citep{Biller2015,Zhou2016} of similar temperatures. Equilibrium condensation models coupled to global circulation models show that refractory species (e.g., CaTiO$_3$, MgSiO$_3$, MnS, Na$_2$S) will also form clouds in the upper atmospheres (typically 10--100 mbar) of hot Jupiters \citep[e.g.,][]{Parmentier2016,Showman2009,Kataria2016}. 
Therefore, our observations suggesting the lack of evidence for a clear atmosphere are fully consistent with the general expectations set by observations and models of brown dwarf and hot Jupiter atmospheres of similar temperatures.

\section{Conclusions} \label{sec:summary}
We have extracted a combined optical transmission spectrum from three transit observations of WASP-4b using \imacs, and use a MultiNest-based retrieval code to test different atmospheric models for the data. We find that a uniform opacity model for the atmosphere is weakly favored over alternatives with hazes, Na, and/or K. In particular, no meaningful evidence for potassium is found despite the elevated radius in the narrow bin centered on the K doublet. Our results, in addition with those of \cite{Huitson17} and \cite{May18}, suggest that no strong signals exist in the optical transit spectrum. Nevertheless, higher quality data may yet reveal evidence for scattering or atomic absorption; for example, \cite{Huitson17} find inconclusive evidence for Na absorption.

We are also able to fit the size and contrast of the star spots occulted by the planet during Transits 2 and 3. Assuming a single spot model, the quality of the data from Transit 3 allows us to tightly constrain the spot size and contrast, which suggest a spot that is much larger and warmer than is typical for spots on the Sun. More complex models which include several spots and faculae could be consistent with the data as well, but are not justified by the evidences. We use this spot model to correct for the wavelength-dependent effect on the transmission spectrum from Transit 3 before averaging the individual nights' spectra. 

Further space-based or larger aperture ground-based observations should be conducted to search for low amplitude signatures of scattering or absorption. However, we note that the presence and strength of a scattering haze is particularly degenerate with the presence of spots and faculae on the star. Since WASP-4 is known to be variable, the stellar photosphere and planet atmosphere should be modeled simultaneously in future analyses of this planet's transmission spectrum.

\acknowledgments
The results reported herein benefited from collaborations and/or information exchange within NASA's Nexus for Exoplanet System Science (NExSS) research coordination network sponsored by NASA's Science Mission Directorate. This paper includes data gathered with the 6.5 meter Magellan-Baade Telescope located at Las Campanas Observatory, Chile. This research has made use of NASA's Astrophysics Data System.

A.B. acknowledges support from the NASA Earth and Space Science Fellowship Program under grant No. 80NSS\-C17K0470. B.R. acknowledges support from the National Science Foundation Graduate Research Fellowship Program under grant No. DGE-1143953.

\software{Astropy \citep{Astropy18}, corner \citep{Foreman-Mackey16}, Matplotlib \citep{Hunter07}, NumPy \citep{Oliphant06}, PyMC \citep{Salvatier16}, PyMultiNest \citep{Buchner14}, SciPy \citep{Jones01}, and SPOTROD \citep{Beky14}.}
 
\newpage
\bibliographystyle{aasjournal}
\bibliography{references}

\appendix

\renewcommand\thefigure{\thesection\arabic{figure}}
\renewcommand\thetable{\thesection\arabic{table}}

\section{Binned light curves}
\setcounter{figure}{0} 
\null
\vfill
\begin{figure}[h]
\centering
\includegraphics[width=0.88\textwidth]{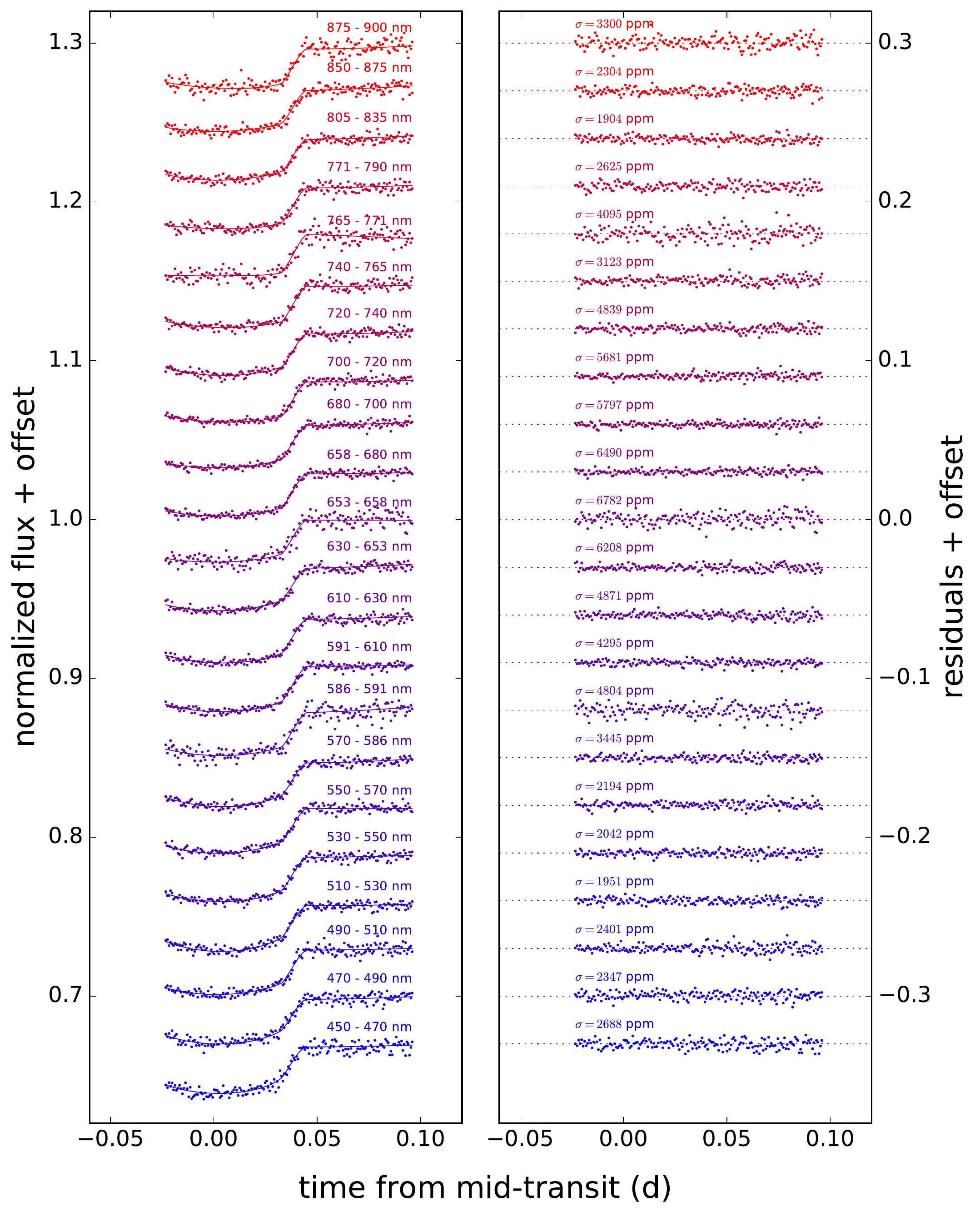}
\caption{Binned light curve fits from the night of 2013-09-24. The models in the left panel include the linear systematics component. \label{fig:130925}}
\end{figure}

\newpage
\null
\vfill
\begin{figure}[h]
\centering
\includegraphics[width=0.88\textwidth]{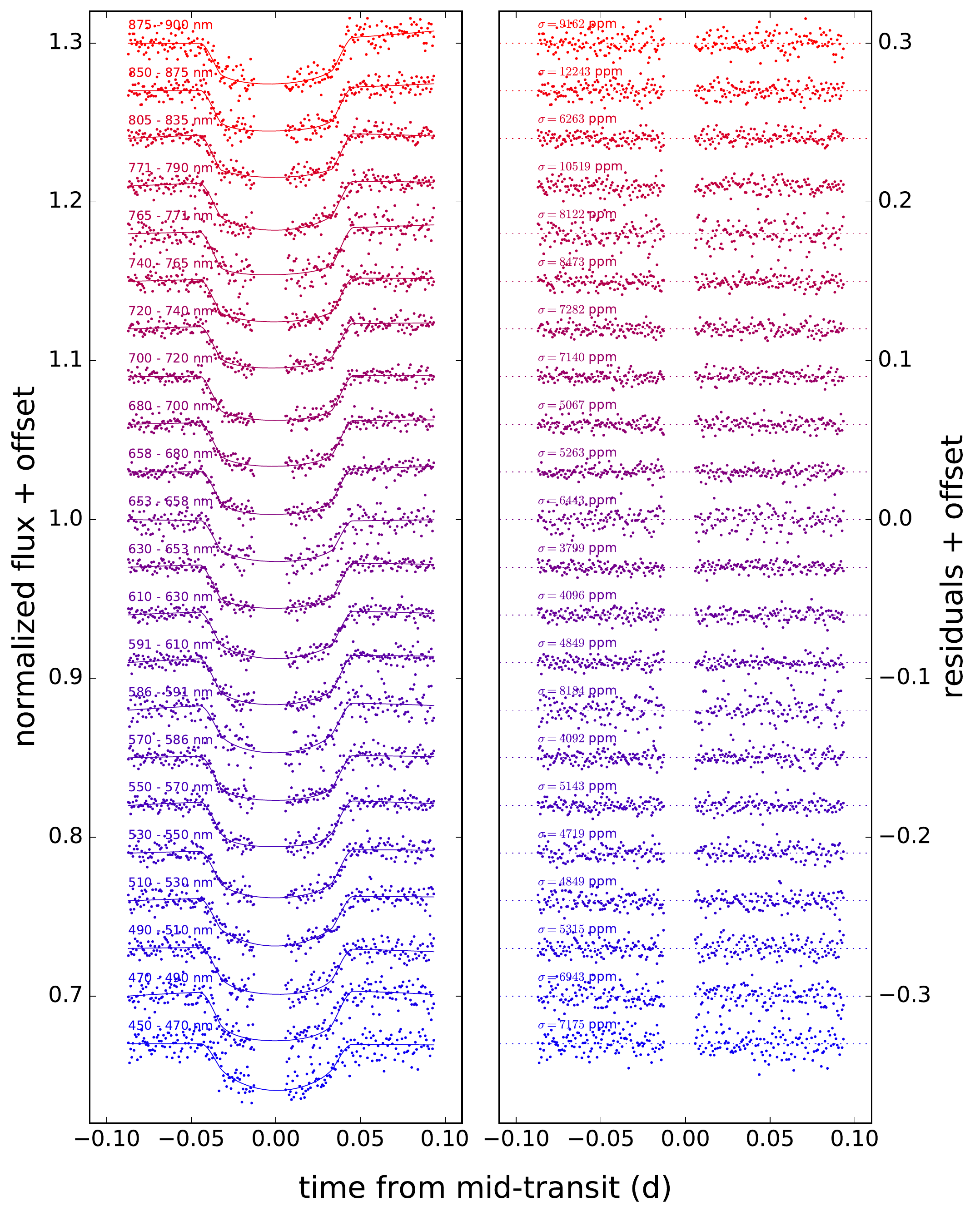}
\caption{Binned light curve fits from the night of 2013-10-17. The models in the left panel include the linear systematics component. A potential spot-crossing event is excluded from the fit.\label{fig:131018}}
\end{figure}

\newpage
\null
\vfill
\begin{figure}[h]
\centering
\includegraphics[width=0.88\textwidth]{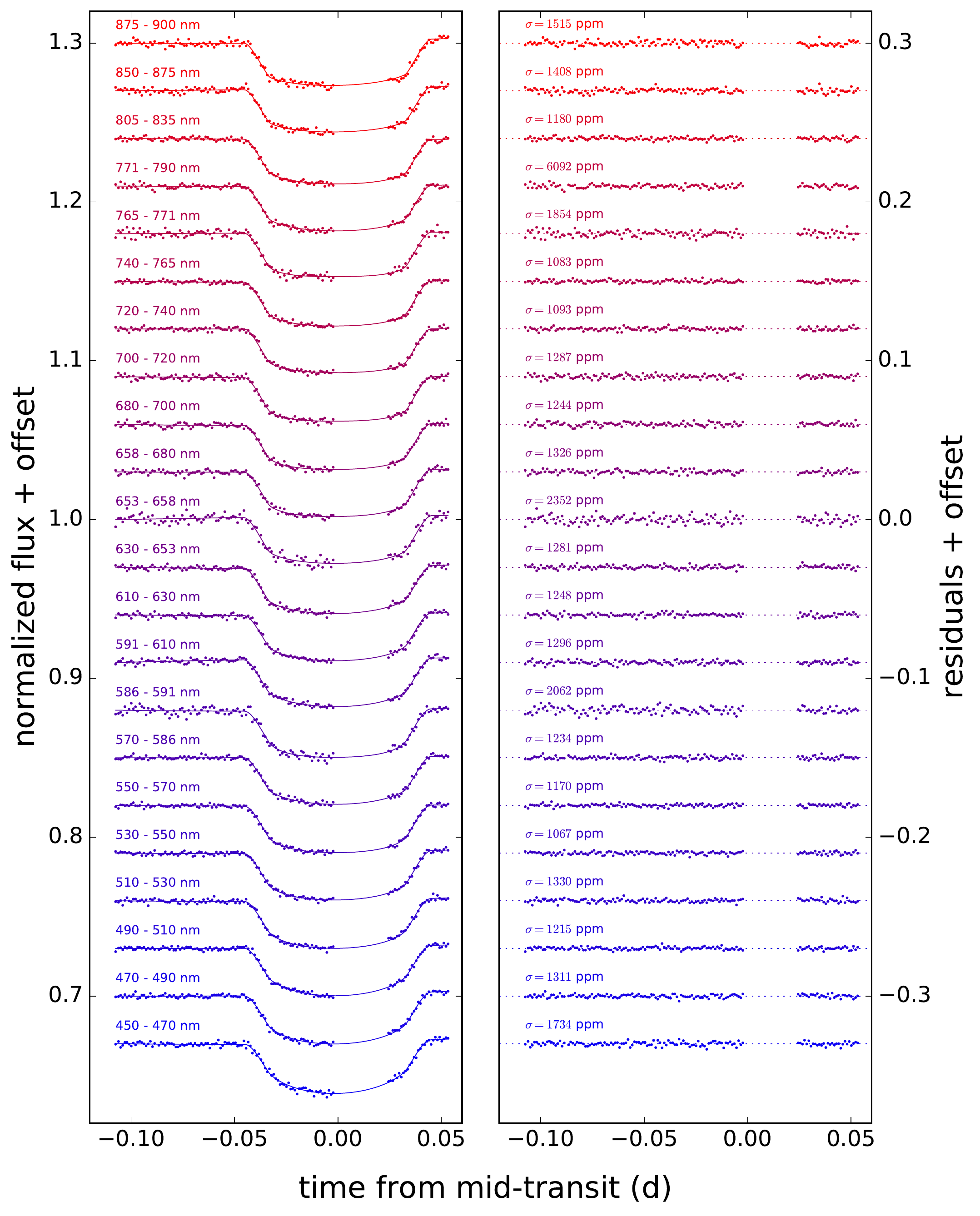}
\caption{Binned light curve fits from the night of 2015-08-14. The models in the left panel include the linear systematics component. The spot-crossing event is excluded.\label{fig:150815}}
\end{figure}

\newpage
\null
\vfill
\begin{figure}[h]
\centering
\includegraphics[width=0.88\textwidth]{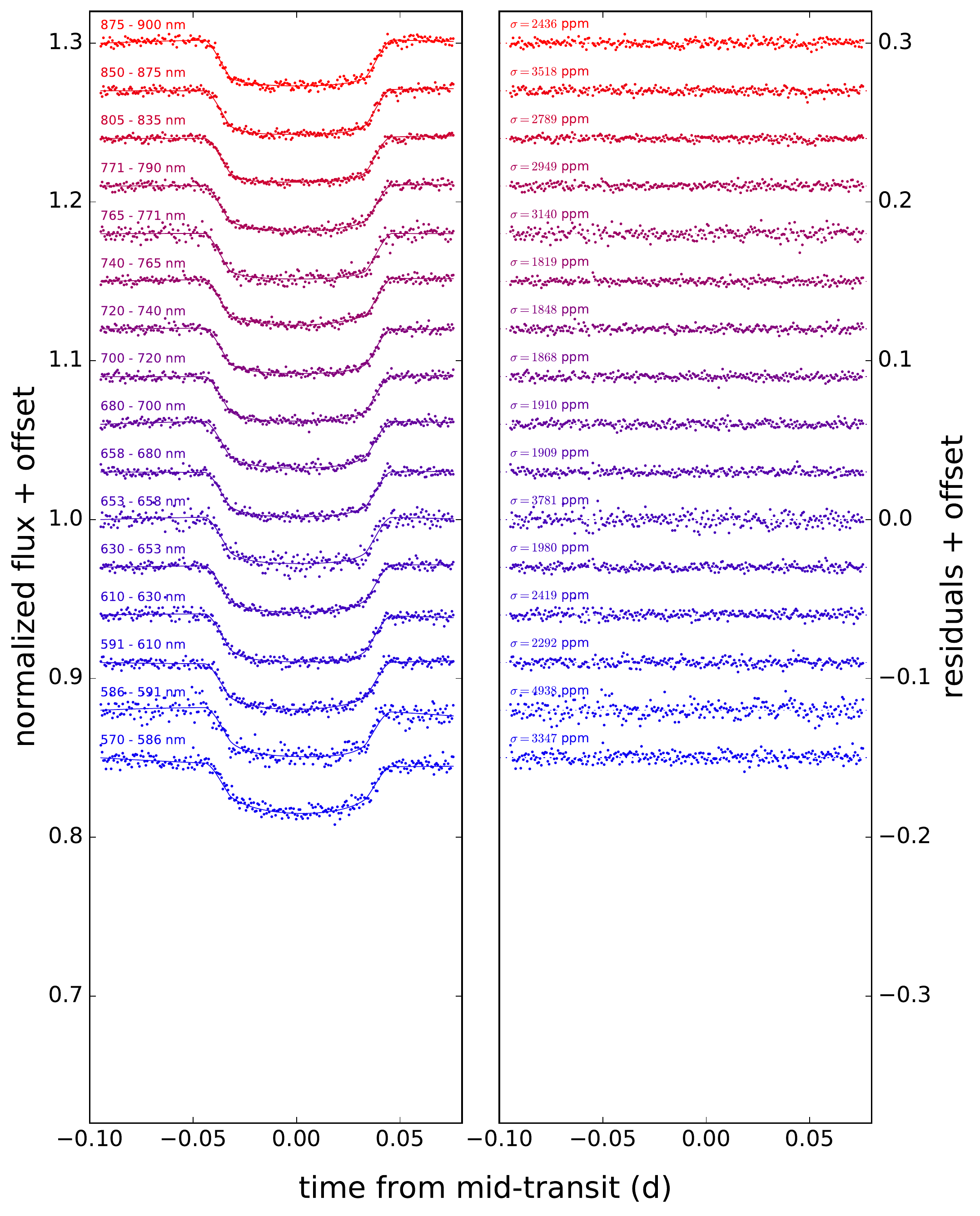}
\caption{Binned light curve fits from the night of 2015-09-26. The models in the left panel include the linear systematics component. Six of the bins are excluded for lack of data due to the narrower filter.\label{fig:150927}}
\end{figure}

\newpage
\begin{figure}[h]
\centering
\includegraphics[width=\textwidth]{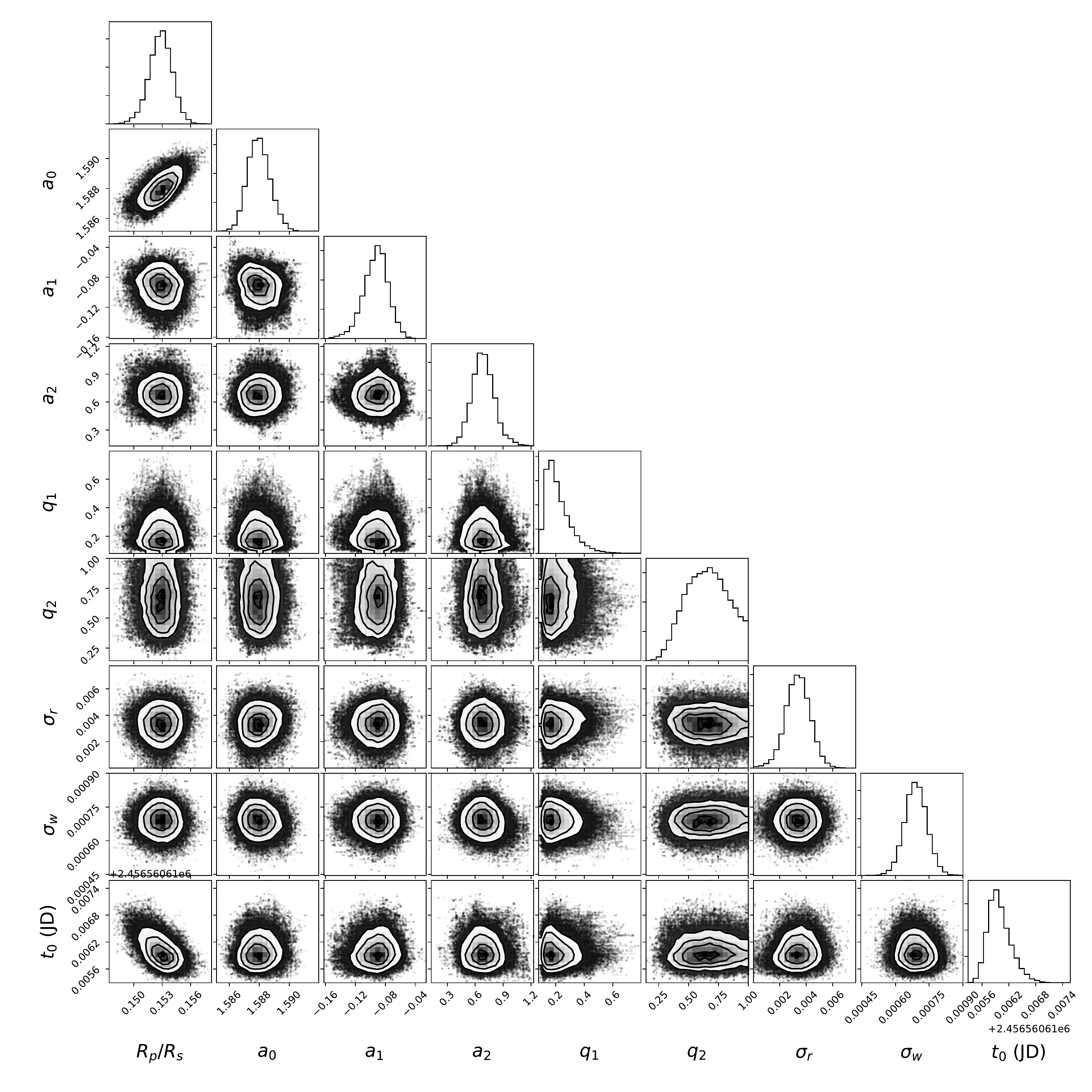}
\caption{Joint posterior distributions for the white light curve model for Transit 1, including the planet-to-star radius ratio ($R_p/R_s$), the correlated and uncorrelated noise parameters ($\sigma_r$, $\sigma_w$), the mid-transit time ($t_0$), three parameters for the quadratic polynomial systematics model ($a_0$, $a_1$, $a_2$), and two parameters for the quadratic limb-darkening law ($q_1$, $q_2$). The light curve of this transit is incomplete, so we exclude its transit spectrum from our combined result. \label{fig:corner_tr1}}
\end{figure}

\newpage
\begin{figure}[h]
\centering
\includegraphics[width=\textwidth]{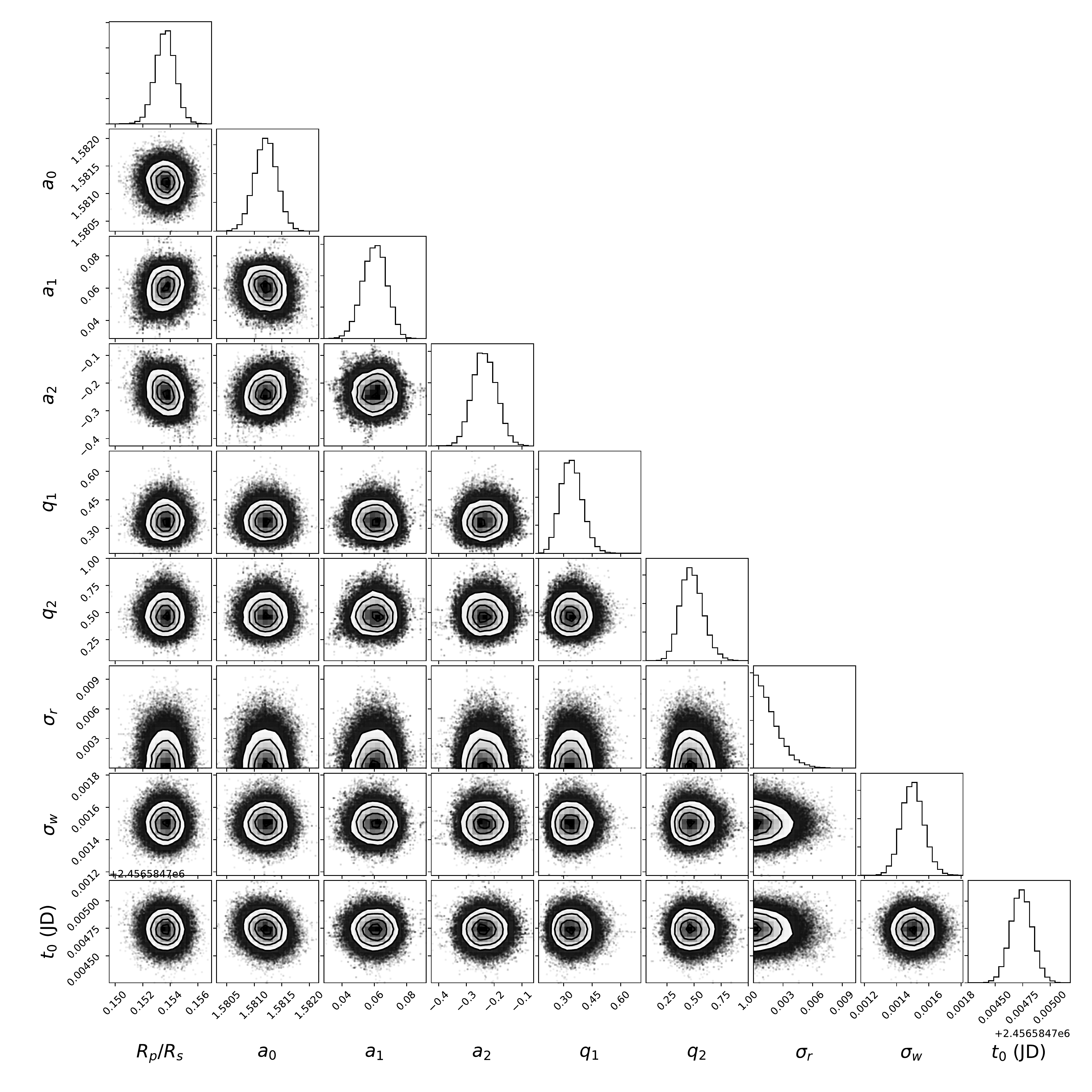}
\caption{Joint posterior distributions for the white light curve model for Transit 2.\label{fig:corner_tr2}}
\end{figure}

\newpage
\begin{figure}[h]
\centering
\includegraphics[width=\textwidth]{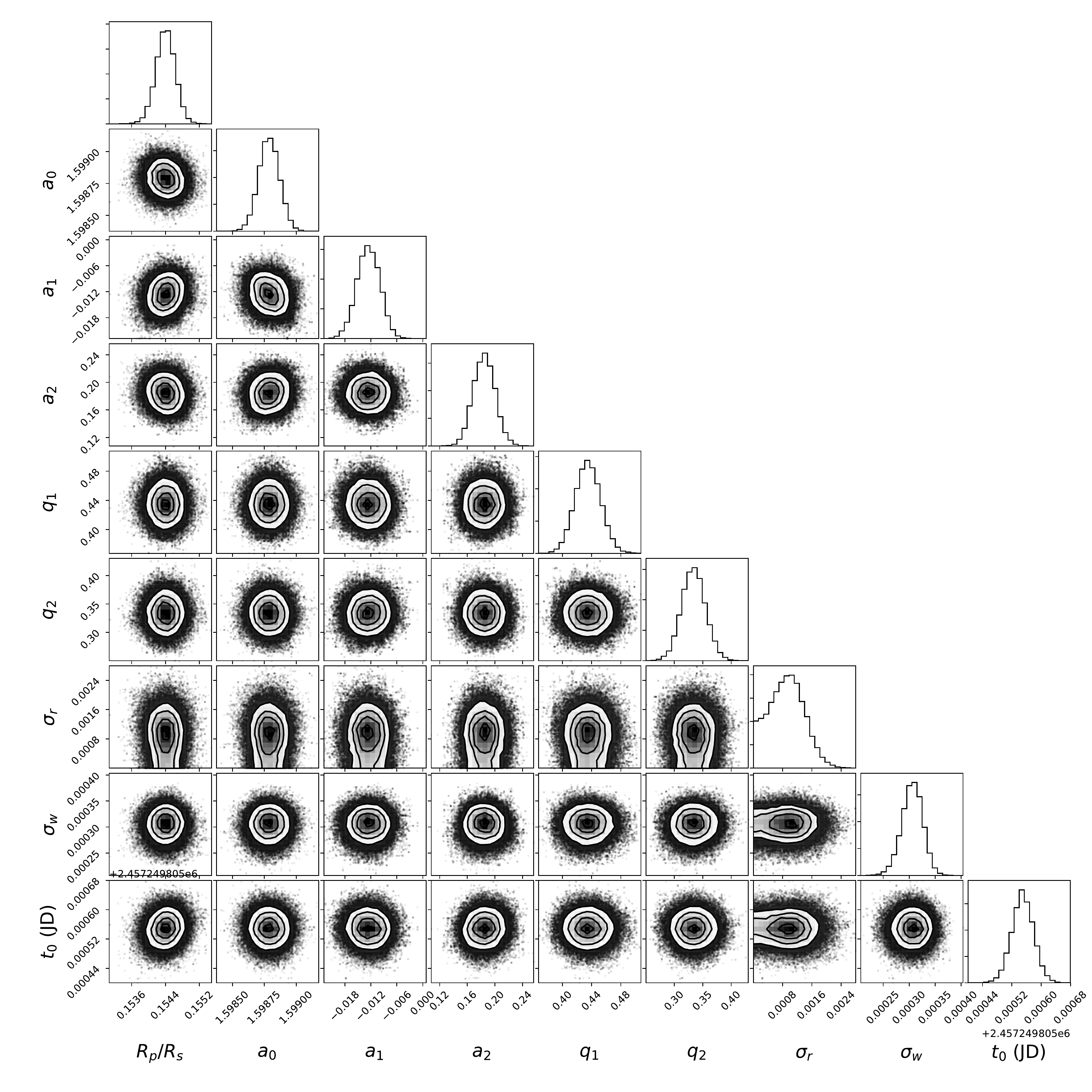}
\caption{Joint posterior distributions for the white light curve model for Transit 3.\label{fig:corner_tr3}}
\end{figure}

\newpage
\begin{figure}[h]
\centering
\includegraphics[width=\textwidth]{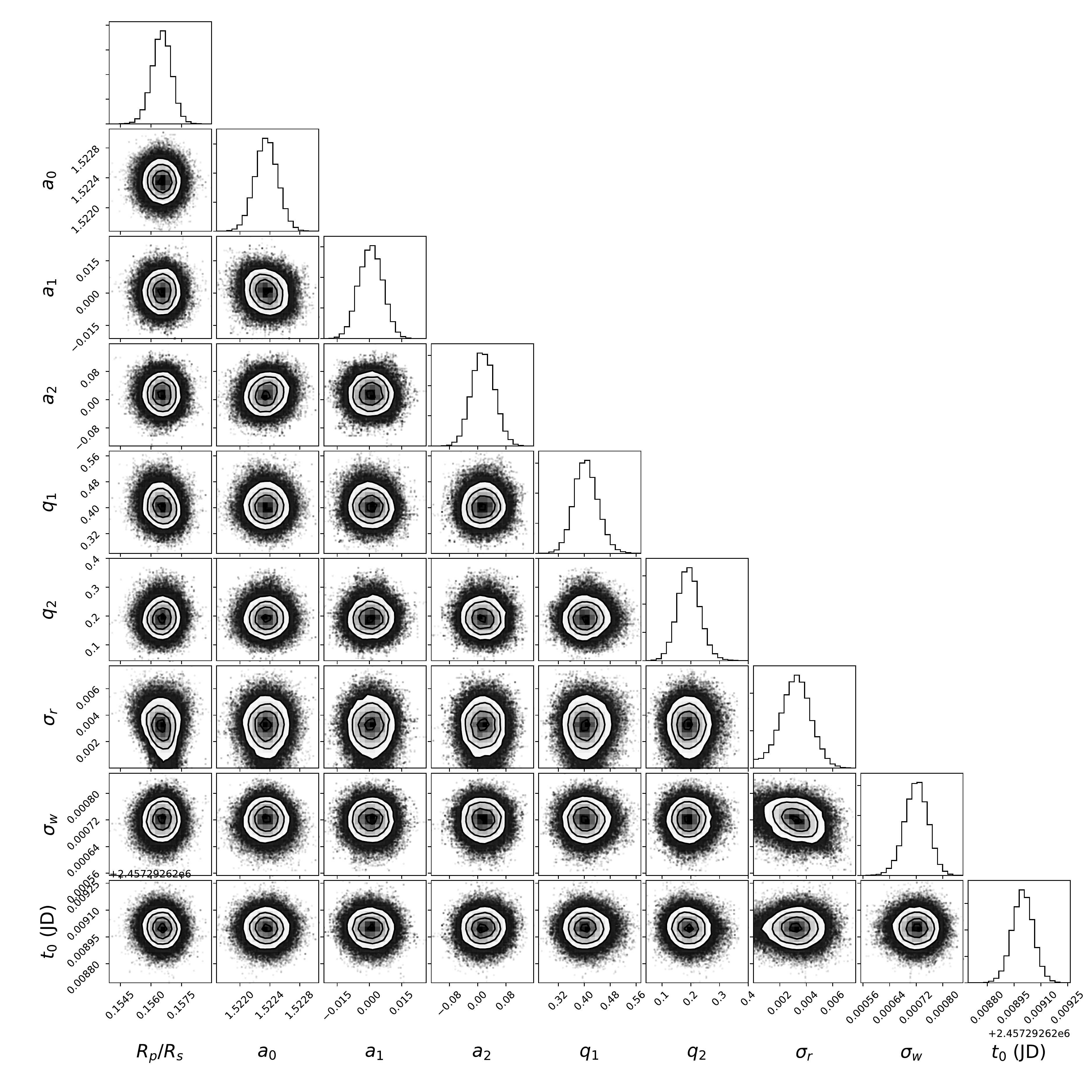}
\caption{Joint posterior distributions for the white light curve model for Transit 4.\label{fig:corner_tr4}}
\end{figure}

\newpage
\begin{figure}[p]
\centering
\includegraphics[width=0.4\textwidth]{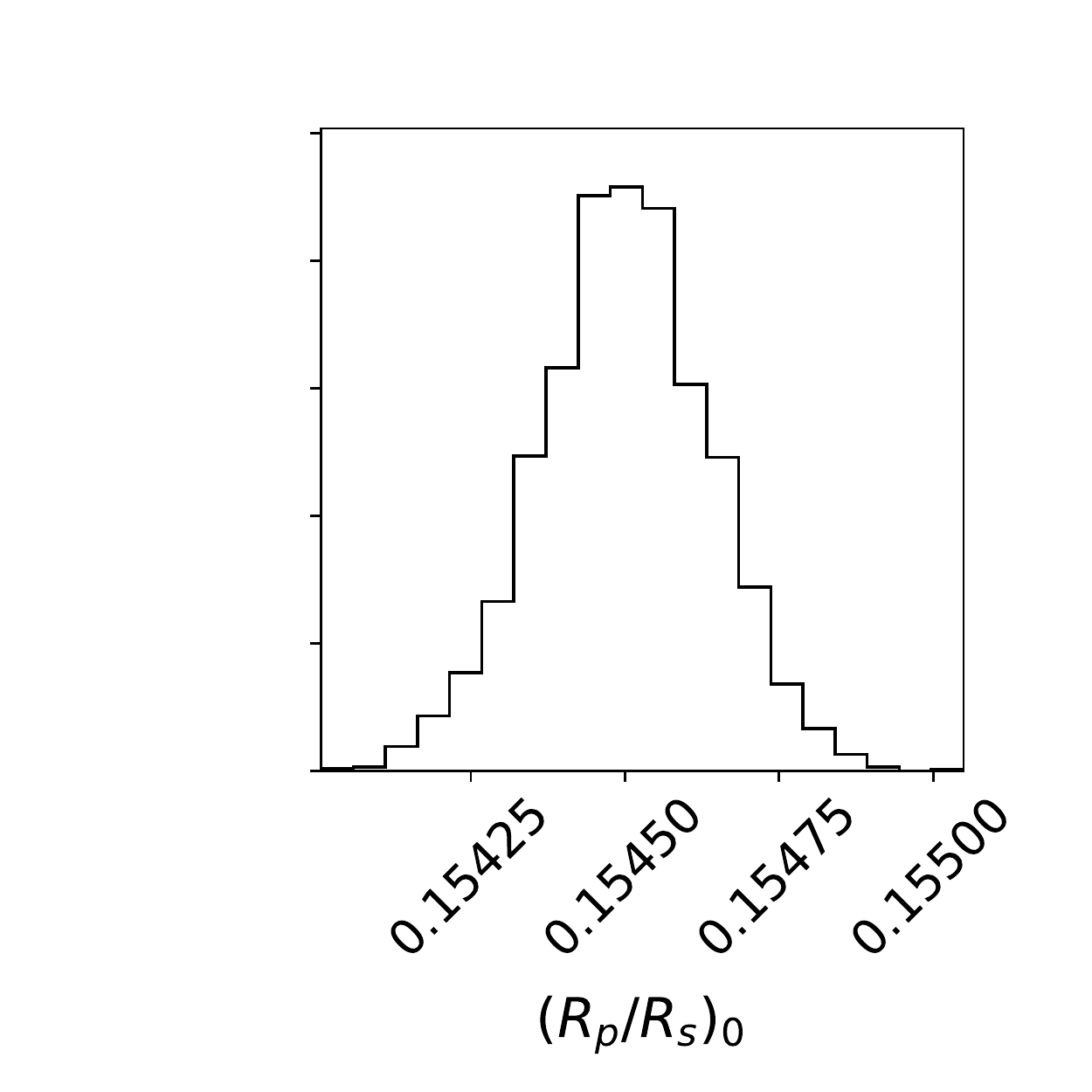}
\includegraphics[width=0.4\textwidth]{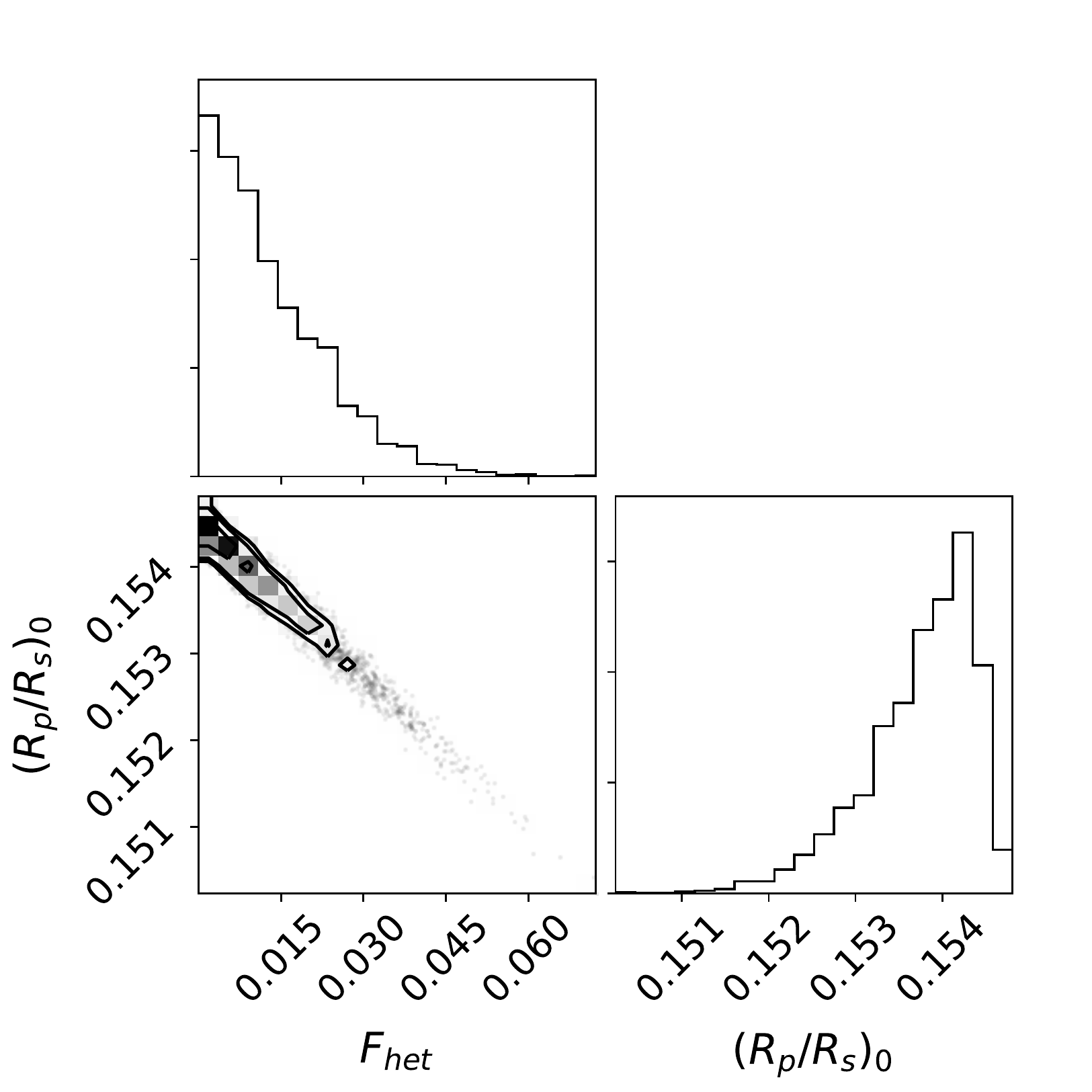}
\includegraphics[width=0.8\textwidth]{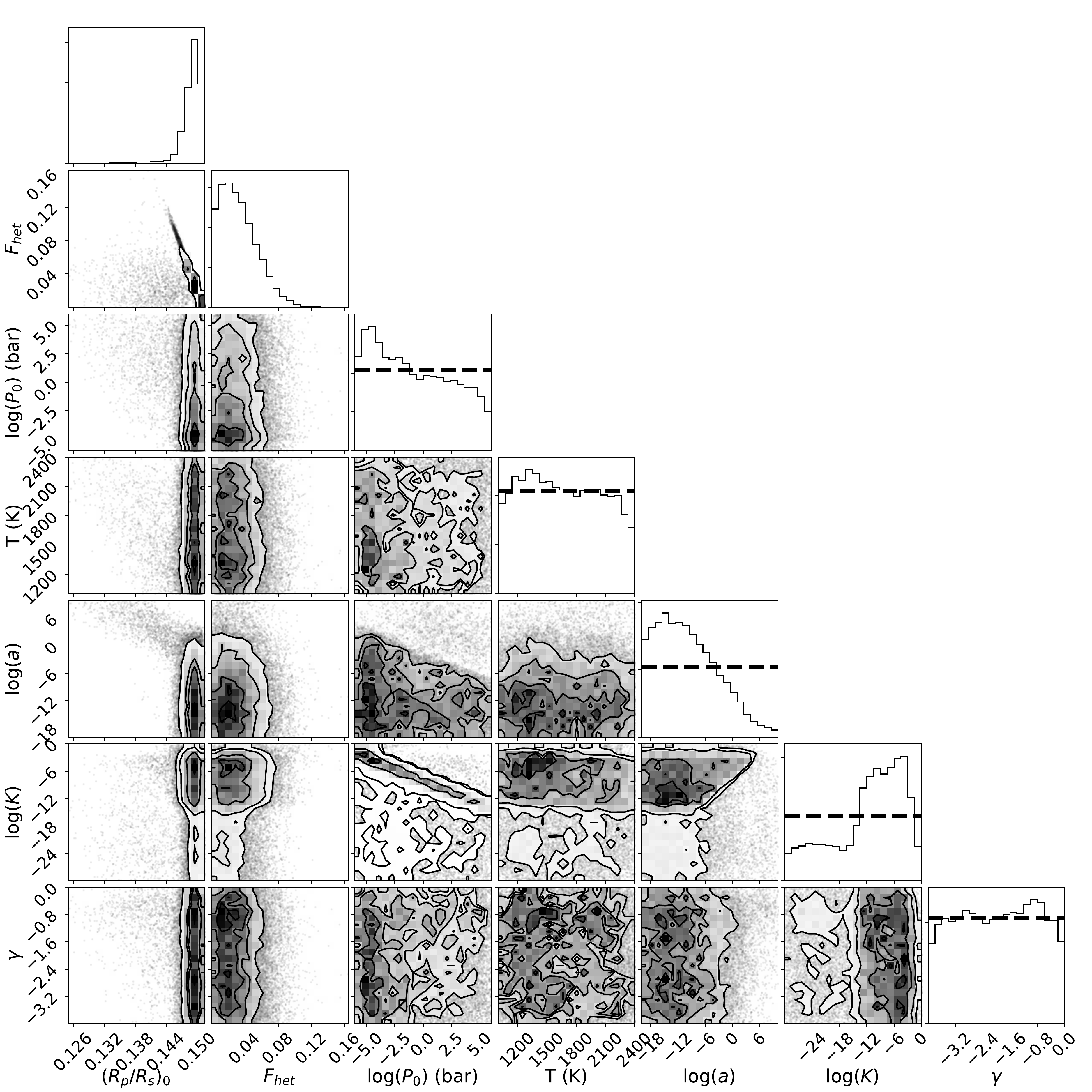}
\caption{Joint posterior distributions for a subset of the atmospheric retrieval models shown in Figure \ref{fig:retrieval}. The parameters and prior distributions are described in Table \ref{tab:priors}. (Top) Models for a uniform opacity atmosphere with (right) and without (left) the contamination features from the photosphere. Here the posterior distributions have converged, and the degeneracy between the planet's radius and the spot covering fraction is apparent. (Bottom) Model for a hazy atmosphere with K absorption and contamination from the photosphere. Some of the parameters do not converge enough from their priors to yield meaningful constraints. In these cases, the prior distributions are marked with a dashed line. \label{fig:corner1}}
\end{figure}

\newpage
\clearpage

\section{Tabulated transmission spectra}
\setcounter{table}{0} 

\begin{deluxetable}{lrlllll}[h]
\centering
\tablecaption{Data for the combined and individual transmission spectra shown in Figures \ref{fig:specs_all} and \ref{fig:spec_combined} with $1\sigma$ uncertainties. The second column is the weighted mean of the mean-subtracted values from Transits 2-4, where the Transit 3 values have first been corrected for the presence of an occulted spot. The last column is the multiplicative effect on the fitted radii for Transit 3 due to the occulted spot, where the true radius $R_{p,0} = R_p/\epsilon^{1/2}$. \label{tab:data}}
\tablehead{\colhead{Bin (nm)}&\colhead{$\Delta R_p/R_s$}&\colhead{Transit 1}&\colhead{Transit 2}&\colhead{Transit 3}&\colhead{Transit 4}&\colhead{$\epsilon^{1/2}$}} 
\startdata
450.0 - 470.0     & -0.0015 $\pm$ 0.0008 & 0.1525 $\pm$ 0.0031 & 0.1430 $\pm$ 0.0028\tddg & 0.1531 $\pm$ 0.0008 &                     & 1.0121 \\
470.0 - 490.0     &  0.0004 $\pm$ 0.0007 & 0.1508 $\pm$ 0.0026 & 0.1583 $\pm$ 0.0045 & 0.1548 $\pm$ 0.0007 &                     & 1.0113 \\
490.0 - 510.0     & -0.0015 $\pm$ 0.0007 & 0.1528 $\pm$ 0.0024 & 0.1514 $\pm$ 0.0033 & 0.1531 $\pm$ 0.0007 &                     & 1.0116 \\
510.0 - 530.0     & -0.0007 $\pm$ 0.0008 & 0.1543 $\pm$ 0.0020 & 0.1533 $\pm$ 0.0029 & 0.1538 $\pm$ 0.0008 &                     & 1.0118 \\
530.0 - 550.0     &  0.0007 $\pm$ 0.0006 & 0.1547 $\pm$ 0.0018 & 0.1550 $\pm$ 0.0029 & 0.1551 $\pm$ 0.0006 &                     & 1.0105 \\
550.0 - 570.0     &  0.0001 $\pm$ 0.0005 & 0.1549 $\pm$ 0.0019 & 0.1530 $\pm$ 0.0021 & 0.1544 $\pm$ 0.0005 &                     & 1.0097 \\
570.0 - 586.8     &  0.0002 $\pm$ 0.0006 & 0.1528 $\pm$ 0.0018 & 0.1520 $\pm$ 0.0025 & 0.1551 $\pm$ 0.0007 & 0.1543 $\pm$ 0.0016 & 1.0092 \\
586.8 - 591.8\tdg & -0.0000 $\pm$ 0.0011 & 0.1529 $\pm$ 0.0038 & 0.1532 $\pm$ 0.0046 & 0.1543 $\pm$ 0.0014 & 0.1569 $\pm$ 0.0020 & 1.0096 \\
591.8 - 610.0     &  0.0015 $\pm$ 0.0006 & 0.1582 $\pm$ 0.0020 & 0.1574 $\pm$ 0.0021 & 0.1566 $\pm$ 0.0007 & 0.1545 $\pm$ 0.0013 & 1.0089 \\
610.0 - 630.0     &  0.0002 $\pm$ 0.0006 & 0.1550 $\pm$ 0.0019 & 0.1552 $\pm$ 0.0023 & 0.1533 $\pm$ 0.0008 & 0.1591 $\pm$ 0.0011 & 1.0089 \\
630.0 - 653.8     &  0.0007 $\pm$ 0.0005 & 0.1492 $\pm$ 0.0022 & 0.1547 $\pm$ 0.0020 & 0.1546 $\pm$ 0.0007 & 0.1579 $\pm$ 0.0010 & 1.0085 \\
653.8 - 658.8\tdg &  0.0010 $\pm$ 0.0012 & 0.1524 $\pm$ 0.0040 & 0.1442 $\pm$ 0.0033 & 0.1576 $\pm$ 0.0017 & 0.1581 $\pm$ 0.0019 & 1.0076 \\
658.8 - 680.0     & -0.0008 $\pm$ 0.0006 & 0.1523 $\pm$ 0.0015 & 0.1496 $\pm$ 0.0022 & 0.1549 $\pm$ 0.0008 & 0.1546 $\pm$ 0.0010 & 1.0082 \\
680.0 - 700.0     &  0.0005 $\pm$ 0.0006 & 0.1516 $\pm$ 0.0018 & 0.1543 $\pm$ 0.0022 & 0.1535 $\pm$ 0.0008 & 0.1585 $\pm$ 0.0008 & 1.0079 \\
700.0 - 720.0     & -0.0005 $\pm$ 0.0006 & 0.1537 $\pm$ 0.0019 & 0.1546 $\pm$ 0.0021 & 0.1539 $\pm$ 0.0008 & 0.1556 $\pm$ 0.0009 & 1.0078 \\
720.0 - 740.0     & -0.0002 $\pm$ 0.0005 & 0.1522 $\pm$ 0.0022 & 0.1514 $\pm$ 0.0024 & 0.1537 $\pm$ 0.0006 & 0.1572 $\pm$ 0.0008 & 1.0076 \\
740.0 - 765.0     &  0.0002 $\pm$ 0.0005 & 0.1559 $\pm$ 0.0029 & 0.1526 $\pm$ 0.0028 & 0.1525 $\pm$ 0.0007 & 0.1588 $\pm$ 0.0007 & 1.0074 \\
765.0 - 771.0\tdg &  0.0017 $\pm$ 0.0009 & 0.1591 $\pm$ 0.0049 & 0.1577 $\pm$ 0.0045 & 0.1548 $\pm$ 0.0011 & 0.1601 $\pm$ 0.0016 & 1.0073 \\
771.0 - 790.0     &  0.0001 $\pm$ 0.0006 & 0.1529 $\pm$ 0.0026 & 0.1571 $\pm$ 0.0031 & 0.1536 $\pm$ 0.0008 & 0.1573 $\pm$ 0.0009 & 1.0071 \\
805.0 - 835.0     & -0.0004 $\pm$ 0.0006 & 0.1475 $\pm$ 0.0020 & 0.1553 $\pm$ 0.0026 & 0.1539 $\pm$ 0.0007 & 0.1555 $\pm$ 0.0010 & 1.0070 \\
850.0 - 875.0     & -0.0014 $\pm$ 0.0007 & 0.1482 $\pm$ 0.0026 & 0.1503 $\pm$ 0.0032 & 0.1527 $\pm$ 0.0011 & 0.1556 $\pm$ 0.0010 & 1.0066 \\
875.0 - 900.0     & -0.0001 $\pm$ 0.0010 & 0.1530 $\pm$ 0.0040 & 0.1495 $\pm$ 0.0040 & 0.1539 $\pm$ 0.0010 & 0.1608 $\pm$ 0.0011\tddg& 1.0064 \\
\enddata
\tablenotetext{\dagger}{Denotes bins centered on Na, H$\alpha$, and K absorption lines.}
\tablenotetext{\ddagger}{These outliers are excluded from the analysis; see Section \ref{sec:excluded_bins}}
\end{deluxetable}

\end{document}